\documentclass[11pt,letterpaper]{article}
\usepackage[margin=1in]{geometry}
\usepackage[T1]{fontenc}
\usepackage[utf8]{inputenc}
\usepackage{lmodern}
\usepackage{microtype}
\usepackage{cite}
\usepackage{amsmath,amssymb,amsthm}
\usepackage{graphicx,booktabs,array,url}
\usepackage{xcolor,tikz}
\usetikzlibrary{arrows.meta,positioning}
\usepackage{seqsplit}
\usepackage[hidelinks]{hyperref}
\hypersetup{
  pdftitle={Polynomial Quantum Recurrent Channels for Matrix Inverse Actions: Deferred-Reset Dynamics and Chebyshev Filtering},
  pdfauthor={Vahid Tavakkoli; Kabeh Mohsenzadegan; Kyandoghere Kyamakya},
  pdfsubject={Research preprint},
  pdfkeywords={polynomial quantum recurrent channels, deferred-reset dynamics, Chebyshev filtering, quantum channels, matrix inverse action}
}

\newtheorem{theorem}{Theorem}

\newtheorem{corollary}{Corollary}
\newcommand{\norm}[1]{\left\lVert #1\right\rVert}
\newcommand{\ket}[1]{\lvert #1\rangle}
\newcommand{\bra}[1]{\langle #1\rvert}
\newcommand{\eps}{\varepsilon}
\newcommand{\tr}{\mathsf T}
\newcommand{\refdoi}[1]{doi:~\href{https://doi.org/#1}{\nolinkurl{#1}}}
\newcommand{\Tr}{\operatorname{Tr}}
\newcommand{\E}{\mathcal E}
\graphicspath{{figures/}}
\title{Polynomial Quantum Recurrent Channels for Matrix Inverse Actions:\\
Deferred-Reset Dynamics and Chebyshev Filtering}
\author{Vahid Tavakkoli$^{1}$, Kabeh Mohsenzadegan$^{1}$, Kyandoghere Kyamakya$^{1,2}$\\[0.6em]
{\small $^{1}$Institute for Smart System Technologies, University of Klagenfurt, Klagenfurt, Austria}\\
{\small $^{2}$Facult\'e Polytechnique, Universit\'e de Kinshasa, Kinshasa, Democratic Republic of the Congo}\\[0.5em]
{\small Corresponding author: Vahid Tavakkoli (\href{mailto:vahid.tavakkoli@aau.at}{vahid.tavakkoli@aau.at})}}
\date{November 2026}

\begin{document}
\maketitle
\begin{abstract}
We develop completely positive, trace-preserving polynomial recurrent channels for quantum solution-state preparation and derive fixed-point and time-varying tracking bounds. Across 2,880 synthetic configurations, resident Chebyshev recurrence improves accuracy-passing snapshot counts in 608 cases, ties in 2,271, and loses in one. A separate 360-scenario cost study shows that fresh recurrence is cheaper in all 184 jointly feasible scenarios (median resident-to-fresh cost ratio: 3.12), while resident recurrence uniquely satisfies accuracy requirements in 74 cases. At a logical noise rate of $10^-4$, neither recurrent method qualifies. Based on exact reduced-density simulations and assumed cost models, our results demonstrate conditional tracking advantages but no general end-to-end speedup or total-cost benefit.
\end{abstract}
\medskip

\section{Introduction}
Polynomial transformations can suppress slowly relaxing residual modes, but in quantum inverse action a useful transformation must also define a physical state update. A nonunitary attractor cannot be implemented by repeatedly applying a closed-system unitary to an arbitrary input state. Its channel must be completely positive and trace-preserving (CPTP), retain a solution-encoding fixed point, and admit an implementation whose oracle, ancilla, reset, and readout costs are explicit. This article addresses that channel-design problem for both static and changing linear systems.

For a nonsingular real matrix $M$ and right-hand side $b$, a quantum linear-system method typically prepares a state related to $M^{-1}b$ rather than writing down the whole inverse matrix~\cite{r1}. Classical recurrent matrix-inversion networks instead evolve an estimate $X$ of the full inverse. The 2019 polynomially weighted recurrent work of Tavakkoli, Chedjou, and Kyamakya \cite{r4} motivates weighting residual modes and studying tracking under perturbations. Its continuous matrix dynamics are not transferred as quantum gates, and the present data register does not store all $N^2$ inverse entries.

Our starting point is an augmented residual $F$, the associated positive contraction $G=F^\dagger F$, and a reference reset state $R$. Augmented solution encoding and residual-reset dissipation have precedent, including Shang's 2026 Lindbladian construction \cite{shang}. The discrete original residual-reset channel is therefore treated as a \emph{reference channel}, not as the primary novelty. We construct two polynomial \emph{recurrent CPTP maps} on top of this interface. Unlike merely computing a Chebyshev approximation to $M^{-1}$, both maps act repeatedly on a resident quantum state and specify how residual weight is discarded and reinjected.

\subsection{Contributions and precise novelty boundary}
The contributions of this paper are as follows.
\begin{enumerate}
\item \textbf{Power-polynomial recurrent channel.} We specify the response $G_m=I-(I-G)^{m+1}$ and the corresponding reset map. We prove exact CPTP validity, preservation of the one-dimensional solution kernel, a unique attracting stationary state, and a convergence certificate. Its \emph{deferred-reset} circuit is not generally the same channel as $m+1$ baseline updates.
\item \textbf{Physical realizations and query accounting.} We give a transformed dense Julia dilation and a construction from controlled calls to the original residual unitary, explicitly accounting for fresh event flags, reference swaps, resets, and environment retention. The increase in convergence per macro update is separated from performance per charged primitive call.
\item \textbf{Signed Chebyshev recurrent channel.} We construct a second CPTP recurrence whose signed polynomial no-event operator suppresses the positive residual spectrum while retaining the solution kernel. We give a shared-call real-Julia QSP realization and prove a conditional $O(\kappa\log(1/\eps))$ versus $\Omega(\kappa^2\log(1/\eps))$ separation \emph{against the defined original recurrent channel} for ideal augmented-state preparation.
\item \textbf{Time-varying matched controls.} We compare a resident Chebyshev channel with the same CPTP map restarted at every snapshot under identical input-oracle ceilings, and with a direct postselected one-shot filter whose expected accepted-state cost is reported separately. We preserve ties, noisy failures, and instances where direct filtering performs better.
\item \textbf{Tracking, readout, and falsifiable resource evidence.} We derive moving-target and finite-implementation-error bounds, extract signed solution components from reference coherences, and compare primitive calls, compiled gates, confidence-dependent readout, logical noise, and synthetic native-gate noise. The study retains failures and noncompetitive outcomes instead of translating simulator convergence into a quantum speedup claim.
\item \textbf{Held-out total-cost frontier.} We assess whether persistent-state tracking reduces complete modeled work after charging destructive-readout copies, trajectory replay, preparations, oracle calls, resets, and input compilation. We select filter degree and update count using validation seeds only and report held-out feasible, infeasible, and higher-cost regions without converting an accuracy gain into a speedup claim.
\end{enumerate}
The elementary power function, Chebyshev polynomials, LCU, eigenstate filtering, QSP, and the augmented residual itself are established ingredients; the proposed contribution is their \emph{specific reset-channel dynamics, realizations, and analysis}. Nor do we claim that these results surpass direct one-shot polynomial filtering, leading quantum linear-system solvers, classical dense solvers, or calibrated quantum hardware. The auxiliary snapshot Chebyshev--LCU inverse-action construction and the continuous classical motivating equations are provided in the accompanying supplementary material to keep the recurrent-channel contribution in focus.

\subsection{Organization}
Section~II positions the work relative to quantum linear-system algorithms and classical recurrent inversion. Sections III--V specify the access model, reference channel, and two polynomial recurrent constructions. The subsequent sections establish tracking and readout bounds, describe the evaluation protocol, and report the numerical and cost-frontier results. The discussion closes by separating genuine channel contributions from established polynomial-filtering techniques.

\section{Related Work and Positioning}
\subsection{Quantum inverse action and polynomial transformations}
Harrow--Hassidim--Lloyd (HHL) established the state-output formulation of quantum linear systems \cite{r1}. Later LCU methods improve precision dependence, and quantum singular value transformation (QSVT) supplies a general framework for transforming block-encoded spectra \cite{r2,r3}. Polynomial eigenstate filtering provides related kernel-selective constructions \cite{filter}. The auxiliary snapshot reciprocal construction uses these established spectral techniques under dense classical access; it is neither the principal novelty of this paper nor an efficient sparse-oracle loader.

Recent functional comparisons emphasize whole-algorithm costs \cite{r5,r6}. Applications to communications and power flow expose the cost of state preparation, condition numbers and classical output requirements \cite{r7,r8,r9}. Sparse loading, general state preparation and synthesis studies explain why logarithmic data-register width alone does not imply logarithmic work \cite{r10,r11,r12,r13}. Near-term HHL bottlenecks and error-mitigation studies also motivate separating an ideal channel, synthetic native noise and calibrated device performance \cite{r14,r15,r16}.

\subsection{Classical recurrent inversion}
The 2019 polynomially weighted recurrent framework is the direct motivation \cite{r4}. Subsequent matrix-flow methods study varying parameters, finite or predefined convergence times, and robustness \cite{r17,r18,r19,r20,r21,r22}. Other approaches develop noise suppression, variable gains, fuzzy control and event-triggered updates \cite{r23,r24,r25,r26,r27}. Generalized-inverse and complex-matrix extensions provide broader classical context \cite{r28,r29,r30,r31,r32,r33,r34}. These works concern classical recurrent states and their own dynamical assumptions. Their rates do not prove a rate for a quantum channel. We use $m$ for polynomial degree and $n_r$ for data-register qubits throughout to avoid conflating degree with storage.

\subsection{Dissipative comparison and scope of novelty}
For residual $F$, Shang's generator has the form
\begin{equation}\label{eq:generator}
\mathcal L(\rho)=\Tr(F^\dagger F\rho)R-\tfrac12\{F^\dagger F,\rho\},\qquad R=\ket r\bra r.
\end{equation}
Its solution encoding and reset structure are established prior work \cite{shang}. The present baseline is the finite collision channel $C\rho C+\Tr(G\rho)R$, where $C=\sqrt{I-G}$. It has the same kernel objective but is not an exact finite-time Lindblad propagator. Its polynomial version is also generally different from applying the baseline $m+1$ times, because a reset is postponed rather than restarting the baseline recurrence inside the macro update.

Our proposed objects are the power-polynomial and signed Chebyshev \emph{recurrent CPTP maps}, with their deferred-reset implementations, tracking and error bounds, and coherent reference readout. The base residual-reset map and the corresponding continuous dissipative generator are comparison constructions; Chebyshev filtering and QSP are not claimed as new polynomial techniques. The numerical Shang comparator implements the stated continuous generator on identical inputs, rather than the fully optimized end-to-end implementation of Shang's algorithm. No gate-count speedup ratio between those solvers is inferred.

\begin{table*}[!tbp]\small\centering
\caption{Novelty audit: established building blocks are distinguished from the two proposed recurrent channels and their proved scope.}
\label{tab:novelty}
\setlength{\tabcolsep}{4.5pt}\begin{tabular}{>{\raggedright\arraybackslash}p{.18\textwidth}>{\raggedright\arraybackslash}p{.245\textwidth}>{\raggedright\arraybackslash}p{.29\textwidth}>{\raggedright\arraybackslash}p{.18\textwidth}}\toprule
Construction & Prior foundation & Specific contribution of this paper & Not established here \\\midrule
Power-polynomial \emph{recurrent channel} & Augmented residual and original reset dynamics \cite{shang}; polynomial mode weighting \cite{r4} & CPTP map with $G_m=I-(I-G)^{m+1}$, deferred-reset realization, fixed-point proof and charged-call audit & General primitive-query improvement or hardware speedup \\
Signed Chebyshev \emph{recurrent channel} & Polynomial eigenstate filtering, QSP/QSVT \cite{r3,filter} & Signed no-event CPTP dynamics, resident-state update, real-Julia reset circuit, conditional separation from original recurrence & Novelty of Chebyshev filtering itself or leading QLSP advantage \\
Reference residual-reset and Shang comparison & Augmented solution-encoding dissipation \cite{shang} & Precisely specified discrete baseline and matched normalized trajectories & Independent invention of residual-reset dissipation \\
Snapshot Chebyshev--LCU action & Reciprocal approximation and LCU \cite{r2,r3} & Auxiliary reproducible operator certificate under dense access & Novel QLSP oracle or end-to-end speedup \\
Classical recurrent motivation & Full-inverse matrix dynamics \cite{r4} & Explicit distinction between classical matrix ODEs and resident-state quantum channels & Literal quantization of the classical matrix flow \\
\bottomrule\end{tabular}
\end{table*}

Recent optimal-query linear-system algorithms provide stronger general-purpose baselines than an unfiltered residual-reset recurrence. Low and Su obtain optimal initial-state-preparation queries under block encoding~\cite{lowsu2026}; Bravo-Prieto, Harrow, and Kothari establish optimal sparse-access query bounds~\cite{bravo2026}. These works assume different access models and are not directly matched to our small dense channel experiments. Accordingly, our cost frontier tests \emph{resident versus restarted operation of the same Chebyshev-channel family}, with direct postselected filtering as a control, not a speedup relative to the leading quantum linear-system algorithms.

\section{Problem and Access Model}
Let $M\in\mathbb R^{N\times N}$ be nonsingular, $b\ne0$, and $x=M^{-1}b$. For real square matrices, $MM^{\tr}\succ0$ is equivalent to nonsingularity. It does not bound the smallest singular value away from zero across a sequence. Write $M=U\Sigma V^{\tr}$ and $0<a\le\sigma_j\le b_s$, with $\kappa=b_s/a$. The auxiliary snapshot Chebyshev formulas denote this spectral upper endpoint by $b=b_s$, which is distinct from the right-hand-side vector. The current experiments use real square matrices. Rectangular pseudoinversion is outside their scope.

For the recurrence define
\begin{equation}\label{eq:normal}
A=\frac{M}{\norm M_2},\quad \beta=\frac{b}{\norm b_2},\quad \gamma=\frac{1}{\kappa},\quad z=\gamma A^{-1}\beta.
\end{equation}
The augmented data dimension is $D=2^{n_r}$, $n_r=\lceil\log_2(N+1)\rceil$. Solution coordinates occupy the first $N$ levels, $\ket r$ is the reference level, and remaining levels are padding. Let
\begin{equation}\label{eq:F}
F=\frac1{\sqrt{1+\gamma^2}}
\begin{bmatrix} A&-\gamma\beta&0\\0&0&0\\0&0&I_{\mathrm{pad}}\end{bmatrix},\quad G=F^\dagger F.
\end{equation}
Then $0\preceq G\preceq I$ and
\begin{equation}\label{eq:psi}
\ket\psi=\frac{(z,1,0)^{\tr}}{\sqrt{1+\norm z_2^2}},\quad \Pi=\ket\psi\bra\psi,\quad q=\bra r\Pi\ket r\ge\tfrac12.
\end{equation}
Indeed $\gamma\le\norm z_2\le1$. Nonzero residual eigenvalues are at least $g=1/(\kappa^2+1)$: the nonzero singular values of $[A,-\gamma\beta]$ are bounded below by those of $A$, and padding has eigenvalue $1/(1+\gamma^2)$. The kernel is one-dimensional.

\subsection{Input, output and costs}
Dense spectral decomposition and synthesis are classical operations. A generic dense factorization costs $O(N^3)$ arithmetic and $O(N^2)$ memory; dense basis synthesis can cost $O(D^2)$ gates. No efficient sparse oracle, quantum random-access memory, fault tolerance or amplitude amplification is implemented. The dense compiler does not explicitly calculate $M^{-1}$ or load $x$, but its singular vectors expose information equivalent to a nullspace solve. This is a strong preprocessing assumption, not an inverse-free quantum advantage. For unrestricted dense inputs, a classical LU or QR solve on the same data provides a natural end-to-end reference: its acquisition, factorization, and solve costs are not paid by a competing oracle in the present asymptotic query theorem. The circuit's ability to avoid an explicit call to an inverse routine at run time therefore does not establish a computational advantage after the spectral compilation has been charged.

The target is the encoded action for a chosen right-hand side. Preparing every column and recovering every real entry of $M^{-1}$ requires further executions and measurements. For $N=1000$, $n_r=10$ augmented data qubits suffice as a representation. The full circuit additionally uses flags, workspace and possibly retained environments. A 42-qubit device is therefore neither excluded solely by $N=1000$ nor established as useful by this width observation.

\section{Reference Residual-Reset Quantum Channel}\label{sec:baseline}
The following discrete channel fixes the comparison target and the primitive residual-call interface. It is a baseline for evaluating the two polynomial recurrences, not the central claim of new polynomial dynamics.
Define $C=\sqrt{I-G}$ and
\begin{equation}\label{eq:baseline}
\E_0(\rho)=C\rho C+\Tr(G\rho)R.
\end{equation}
The first branch preserves the no-event state, and the second resets residual weight to the known reference. Kraus operators are $C$ and $\ket r\bra j F$ for a complete basis $j$. Their adjoint products sum to $I-G+G=I$, proving complete positivity and trace preservation. Since $F\ket\psi=0$ and $C\ket\psi=\ket\psi$, the target is fixed.

\begin{theorem}[Global attraction]\label{thm:base}
For every initial density matrix, $u_K=1-\Tr(\Pi\rho_K)$ satisfies
\begin{equation}\label{eq:basebound}
u_K\le(1-qg_+)^K u_0\le(1-g/2)^K u_0,
\end{equation}
where $g_+$ is the actual smallest positive eigenvalue of $G$ and $g=1/(\kappa^2+1)$. Thus $\rho_K$ converges to the unique stationary state $\Pi$.
\end{theorem}
\begin{proof}
Because $C\Pi C=\Pi$, $\Tr(\Pi\E_0(\rho))=\Tr(\Pi\rho)+q\Tr(G\rho)$. The inequality $G\succeq g_+(I-\Pi)$ gives $u_{K+1}\le(1-qg_+)u_K$. Iteration proves the bound. A state with unit overlap with a rank-one projector is that projector; the bound forces any stationary state to have this overlap. Purified distance to a pure target is $\sqrt{u_K}$, which proves convergence in trace distance as well.
\end{proof}
The theorem covers mixed, basis and coherent initial states, including ones with zero initial reference population. A reference start has $u_0=1-q\le1/2$. Coherence readout can be undefined at an early iterate with zero reference population; the numerical records preserve that status.

\section{Power-Polynomial Quantum Recurrent Channel}\label{sec:power}
\textbf{First proposed recurrent construction.} The key operation is not a polynomial applied to a final measured solution: it is a polynomially transformed no-event branch of a CPTP map, with residual events collected and reset only once at the macro boundary. Its state persists across updates.
Put $\ell=m+1$ and define
\begin{align}
q_m(g)&=1-(1-g)^{\ell},\label{eq:response}\\
G_m&=q_m(G),\quad C_m=C^{\ell},\label{eq:transform}\\
\E_m(\rho)&=C_m\rho C_m+\Tr(G_m\rho)R.\label{eq:accelerated}
\end{align}
Here ``acceleration means a larger certified contraction per \emph{macro update}. It does not by itself mean fewer charged primitive calls or faster end-to-end computation; the comparison to the baseline is derived below.

\begin{theorem}[Validity, kernel and convergence]\label{thm:accelerated}
For $m\ge0$, \eqref{eq:accelerated} is completely positive and trace-preserving. Its intended solution kernel is unchanged, its unique stationary state is $\Pi$, and
\begin{equation}\label{eq:polybound}
u_K\le(1-q q_m(g_+))^K u_0\le(1-\theta_m)^K u_0,
\end{equation}
where
\begin{equation}\label{eq:theta}
\theta_m=\frac12\left[1-\left(\frac{\kappa^2}{\kappa^2+1}\right)^{m+1}\right].
\end{equation}
\end{theorem}
\begin{proof}
Functional calculus on $0\preceq G\preceq I$ gives $0\preceq G_m\preceq I$ and $C_m^2=I-G_m$. Define $H_m=\sum_{j=0}^{m}(I-G)^j$ and $F_m=F\sqrt{H_m}$. Then $F_m^\dagger F_m=GH_m=G_m$. Kraus operators $C_m$ and $\ket r\bra j F_m$ are complete. Since $H_m\succeq I$ and $q_m(g)=0$ exactly at $g=0$, $\ker F_m=\ker F$. The polynomial is strictly increasing on $[0,1)$; hence the positive spectral gap is exactly $q_m(g_+)$. Repeating the overlap proof of Theorem~\ref{thm:base} proves global attraction and uniqueness.
\end{proof}

\begin{corollary}[Accuracy and charged rate]\label{cor:cost}
To achieve purified distance at most $\eps$, it suffices that
\begin{equation}\label{eq:K}
K\ge\left\lceil\frac{\log(u_0/\eps^2)}{-\log(1-\theta_m)}\right\rceil,
\end{equation}
with zero updates if $u_0\le\eps^2$. The macro bound is
$O([1+(\kappa^2+1)/\ell]\log(1/\eps))$ for fixed initial error. Charging $\ell$ primitive residual calls per macro gives no asymptotic improvement from this bound alone.
\end{corollary}
\begin{proof}
For $0\le g\le1$, $\ell g/(1+\ell g)\le q_m(g)\le\min(1,\ell g)$. The upper bound is Bernoulli's inequality; the lower follows from $(1-g)^{-\ell}\ge1+\ell g$ for $g<1$, with the endpoint by continuity. These bounds and $-\log(1-\theta)\ge\theta$ give the macro estimate. More sharply, convexity gives
\begin{equation}\label{eq:charged}
\tfrac12[1+(1-g)^\ell]\ge(1-g/2)^\ell.
\end{equation}
Thus even the universal certificate exponent per charged call cannot improve through the update-count argument. This compares sufficient bounds, not all realized trajectories.
\end{proof}
The map differs from $\E_0^{\ell}$ because a residual event is retained through the macro and reset once at its end. The solution-sector probability at stationarity remains $\norm z^2/(1+\norm z^2)\in[1/(\kappa^2+1),1/2]$; polynomial weighting does not improve it.

\subsection{An explicit dense dilation}
For a contraction $F_m$, the Julia unitary
\begin{equation}\label{eq:julia}
U_m=\begin{bmatrix}C_m&-F_m^\dagger\\F_m&\sqrt{I-F_mF_m^\dagger}\end{bmatrix}
\end{equation}
is unitary, since $F_m C_m=\sqrt{I-F_mF_m^\dagger}F_m$. Applying it to a zero flag and the data produces the no-event and residual branches. A flag-controlled swap with a fresh $\ket r$ register replaces the residual data by $\ket r$. Tracing the flag and swapped workspace gives exactly \eqref{eq:accelerated}.

The compiler obtains $F_m$ from the singular value decomposition of $F$, replacing each singular value $s$ by $\sqrt{1-(1-s^2)^\ell}$. The target inverse is never used as circuit input. Nonetheless, this dense decomposition already exposes the residual nullspace; its classical cost must be counted. This realization establishes physical validity and small-system implementability, not an efficient matrix-oracle algorithm.

\subsection{Controlled baseline calls and deferred reset}
\begin{figure}[!tbp]\centering
\begin{tikzpicture}[node distance=4mm,font=\scriptsize,>=Latex]
\node[draw,rounded corners,text width=7cm,align=center] (a) {Resident augmented data $\rho$; $\ell=m+1$ fresh zero event flags};
\node[draw,below=of a,text width=7cm,align=center] (b) {Controlled $U_0$ calls: call $j$ enabled only while all earlier flags are zero};
\node[draw,below=of b,text width=7cm,align=center] (c) {Defer the residual event; preserve branches $C^\ell$ and $FC^j$};
\node[draw,below=of c,text width=7cm,align=center] (d) {Flag-controlled swap with fresh reference $\ket r$; discard or retain environment};
\node[draw,below=of d,text width=7cm,align=center] (e) {Same physical data register enters next macro update};
\draw[->](a)--(b);\draw[->](b)--(c);\draw[->](c)--(d);\draw[->](d)--(e);
\end{tikzpicture}
\caption{Realizable polynomial channel using deferred resets. Resetting immediately after each primitive call would implement a different channel. No measurement-driven classical solution correction occurs.}\label{fig:architecture}
\end{figure}
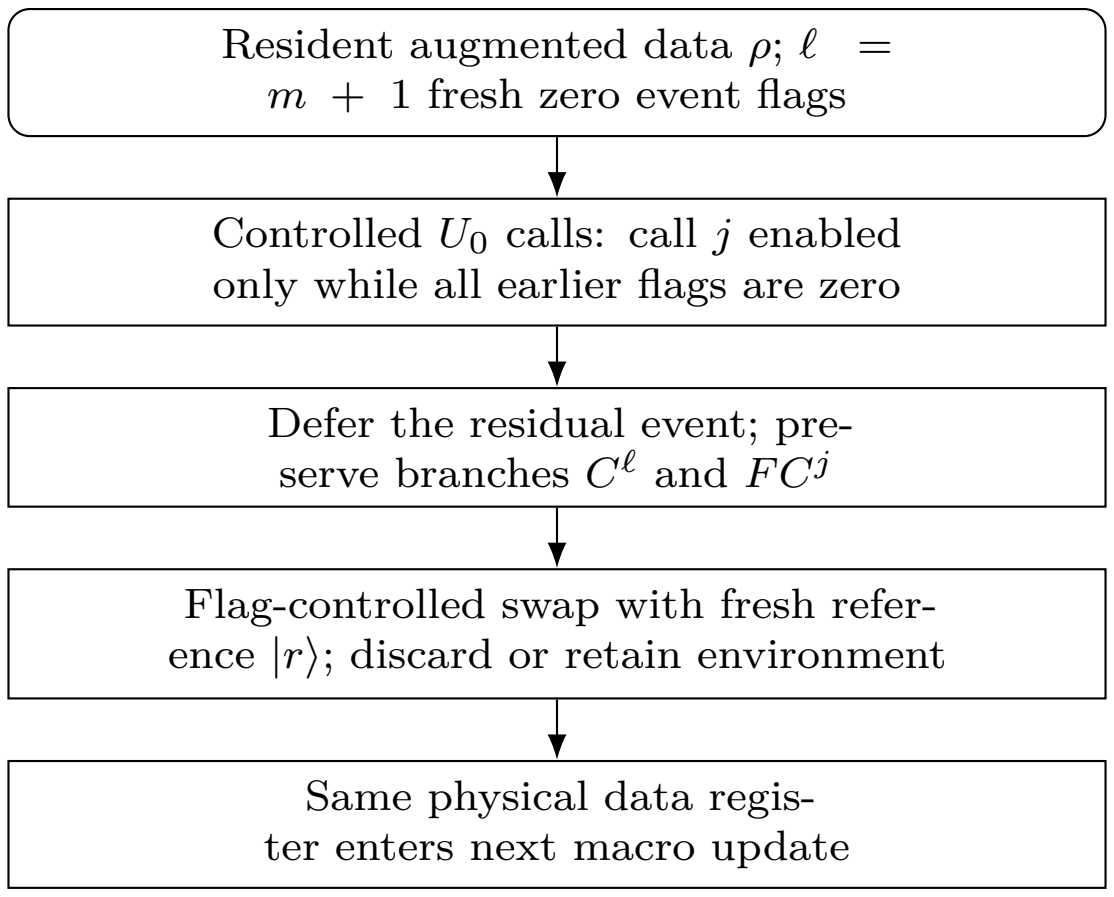

A second implementation uses only the original residual unitary $U_0$, not a new transformed-factor decomposition. Prepare $\ell$ zero flags. Apply $U_0$ with flag $j$ only if every previous flag is zero. The first event at step $j$, starting at zero, has data operator $FC^j$; later residual calls are disabled. After all calls, swap the data to the reference conditioned on the event flags. At most one event flag is set. The branch completeness identity is
\begin{equation}\label{eq:telescope}
\sum_{j=0}^{\ell-1}C^jGC^j=I-C^{2\ell}=G_m.
\end{equation}
The no-event branch is $C^\ell$. Orthogonal flags remove cross terms after tracing, so the circuit implements precisely $\E_m$. Figure~\ref{fig:architecture} summarizes this deferred-reset construction.

The data register persists across updates. On a reset-capable device, the environment is reset and reused between macro updates. With coherent retention, each update receives fresh environment qubits and the global state remains unitary. The reduced data state can converge while the total retained state does not converge to a universal pure state. This resolves the distinction between an entirely quantum circuit and a closed register that cannot erase entropy.

\subsection{Resource accounting}
The dense dilation uses $2n_r+1$ qubits with reset reuse, or $n_r+K(n_r+1)$ with coherent retention. The controlled-call construction uses $2n_r+\ell$ with resets, or $n_r+K(n_r+\ell)$ with retained environments. It charges $K\ell$ residual-unitary calls, $K\ell(\ell-1)/2$ additional control incidences, at most $Kn_r\ell$ controlled swaps, and $(K-1)(n_r+\ell)$ ancillary resets. Multi-controlled synthesis, routing, preparation and measurements are additional costs. Counts reported below come from actual constructed native circuits.

The power-response construction does not supply a query improvement merely through a larger macro gap. The next section develops a distinct bounded Chebyshev response and its explicit real-Julia signal-processing circuit. Its query analysis uses a specified residual-unitary interface, while dense preprocessing and output costs remain separate.


\section{Signed Chebyshev Polynomial Recurrent Channel}\label{sec:cheb}
\textbf{Second proposed recurrent construction.} Unlike the power response, this channel uses a signed no-event polynomial. The polynomial filtering principle is established \cite{filter}; the object analyzed here is the associated \emph{resident-state CPTP recurrence}, its reset circuit, and its conditional query behavior relative to the reference channel.
\subsection{What the power certificate cannot establish}
For the actual reference overlap $q\in(0,1]$, convexity of $x\mapsto x^\ell$ gives
\begin{equation}\label{eq:powerq}
1-q[1-(1-g)^\ell]\ge(1-qg)^\ell.
\end{equation}
Indeed the left side is the convex average of $1$ and $(1-g)^\ell$ with weights $1-q,q$. Thus the power-response certificate per $\ell$ charged calls is no better than the original certificate. This generalizes \eqref{eq:charged}; it does not prove that every compiled implementation or trajectory is slower. The retained equal-call audit and the dense-compiler savings answer different cost questions.

\subsection{Bounded signed response}
Use $s\ge1$ for the recurrent Chebyshev degree, distinct from reciprocal degree $d$ and power degree $m$. Given a certified positive spectral gap $0<g<1$, define
\begin{equation}\label{eq:chebr}
\begin{aligned}
r_s(\lambda)&=\frac{T_s((1+g-2\lambda)/(1-g))}{T_s((1+g)/(1-g))},\\
\eta_s&=\operatorname{sech}(2s\operatorname{artanh}\sqrt g).
\end{aligned}
\end{equation}
Then $r_s(0)=1$, $|r_s(\lambda)|\le1$ on $[0,1]$, and $|r_s(\lambda)|\le\eta_s<1$ on $[g,1]$. For the latter interval, the numerator argument lies in $[-1,1]$. On $[0,g]$ it lies between $1$ and $(1+g)/(1-g)$, where $T_s$ is positive and increasing. The identity $\operatorname{arcosh}[(1+g)/(1-g)]=2\operatorname{artanh}\sqrt g$ gives
\begin{equation}\label{eq:chebeta}
\eta_s\le2e^{-2s\operatorname{artanh}\sqrt g},\quad
s\ge\left\lceil\frac{\operatorname{arcosh}(1/\eta)}{2\operatorname{artanh}\sqrt g}\right\rceil
\end{equation}
as a sufficient degree for a chosen $\eta_s\le\eta<1$. This is established Chebyshev eigenstate filtering \cite{filter}; its use here is not claimed as invention of a new polynomial approximation.

Set $B_s=r_s(G)$ and $H_s=I-B_s^2$. Our signed-filter \emph{recurrent channel} is
\begin{equation}\label{eq:chebchannel}
\E_s^{\mathrm{Ch}}(\rho)=B_s\rho B_s+\Tr(H_s\rho)R.
\end{equation}
Negative eigenvalues of $B_s$ are intentional. Replacing them by positive roots would define a different coherent no-event branch.

\begin{theorem}[Signed-filter channel]\label{thm:cheb}
For a positive contraction $G$ with one-dimensional kernel $\Pi$ and nonzero spectrum in $[g,1]$, with $q=\Tr(\Pi R)>0$, \eqref{eq:chebchannel} is CPTP, preserves that kernel through $H_s$, and has unique stationary state $\Pi$. For every initial density matrix,
\begin{equation}\label{eq:chebrate}
u_K\le[1-q(1-\eta_s^2)]^K u_0.
\end{equation}
If $q\ge1/2$ and $\eta_s\le1/2$, the infidelity contracts by at most $5/8$ per macro.
\end{theorem}
\begin{proof}
The spectral bounds imply $H_s\succeq0$. Kraus operators $B_s$ and $\ket r\bra j\sqrt{H_s}$ have adjoint products summing to $I$. Since $r_s(0)=1$ and $|r_s(\lambda)|<1$ for positive spectral values, $\ker H_s=\ker G$ and $B_s\Pi B_s=\Pi$. As before, the target overlap increases exactly by $q\Tr(H_s\rho)$. The inequality $H_s\succeq(1-\eta_s^2)(I-\Pi)$ proves the contraction, uniqueness and global attraction. No positivity assumption on $B_s$ is required.
\end{proof}

\subsection{A shared-oracle signal-processing circuit}
We specify a stronger interface than an unspecified block encoding: an exact \emph{real Julia dilation} $U_0$ of the input residual $F$, already used by the original recurrence. The gap bound and flag phase rotations are also available. Classical construction of that oracle is excluded only from the query theorem; it is included in the implementation cost study.

If $F=L\Sigma_FV^{\tr}$, changing the flag-zero basis by $V$ and flag-one basis by $L$ decomposes $U_0$ into real rotations
\[
J(x)=\begin{bmatrix}x&-\sqrt{1-x^2}\\\sqrt{1-x^2}&x\end{bmatrix},\qquad x\in\operatorname{spec}(C).
\]
These basis changes are used in the proof, not executed or supplied as extra free oracles. Flag conjugation with $S=\operatorname{diag}(1,i)$ gives $SJ(x)S^\dagger=W_x(x)$, the standard QSP signal matrix. The even polynomial
\begin{equation}\label{eq:evenqsp}
p_{2s}(x)=r_s(1-x^2),\qquad |p_{2s}(x)|\le1\quad(-1\le x\le1)
\end{equation}
is achievable as a QSP quadrature by standard real-polynomial achievability \cite{r3,pyqsp}. Let the top signal block of a phase word $V_\Phi$ have imaginary part $p_{2s}(C)$. Since $U_0$ is real, phase-conjugate branches $V_\Phi$ and $V_\Phi^*$ share the \emph{same forward} input calls. A selector controls only the signs of phase rotations. Preparing and recombining that selector extracts
\begin{equation}\label{eq:quadrature}
\frac{V_\Phi-V_\Phi^*}{2i}
\end{equation}
as its selected block, whose simultaneous flag-zero data block is $r_s(G)$. There are exactly $2s$ calls to $U_0$, not two separately executed phase words. This optimization uses the real-Julia structure and is not asserted for an arbitrary complex block encoding.

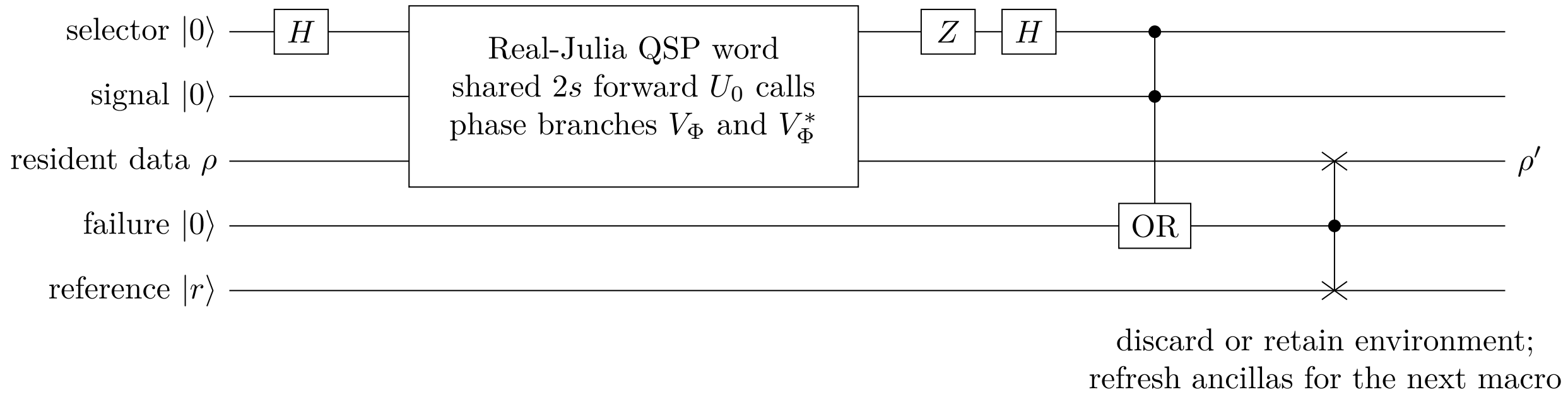
\begin{figure*}[!tbp]\centering
\resizebox{\textwidth}{!}{%
\begin{tikzpicture}[x=1cm,y=.72cm,font=\small]
\foreach \y in {3,2,1,0,-1}{\draw (0,\y)--(14.2,\y);}
\node[left] at (0,3) {selector $\ket0$};\node[left] at (0,2) {signal $\ket0$};
\node[left] at (0,1) {resident data $\rho$};\node[left] at (0,0) {failure $\ket0$};
\node[left] at (0,-1) {reference $\ket r$};
\node[draw,fill=white,minimum width=.6cm,minimum height=.4cm] at (.8,3) {$H$};
\draw[fill=white] (2,.6) rectangle (7,3.4);
\node[align=center] at (4.5,2.1) {Real-Julia QSP word\\shared $2s$ forward $U_0$ calls\\phase branches $V_\Phi$ and $V_\Phi^*$};
\node[draw,fill=white,minimum width=.6cm,minimum height=.4cm] at (8,3) {$Z$};
\node[draw,fill=white,minimum width=.6cm,minimum height=.4cm] at (8.9,3) {$H$};
\draw (10.3,3)--(10.3,0);\fill (10.3,3) circle(2pt);\fill (10.3,2) circle(2pt);
\node[draw,fill=white,minimum width=.8cm,minimum height=.42cm] at (10.3,0) {OR};
\draw (12.3,-1)--(12.3,1);\fill (12.3,0) circle(2pt);
\foreach \y in {1,-1}{\draw (12.16,\y-.14)--(12.44,\y+.14);\draw (12.16,\y+.14)--(12.44,\y-.14);}
\node[right] at (14.2,1) {$\rho'$};
\node[below,align=center] at (12.2,-1.4) {discard or retain environment;\\refresh ancillas for the next macro};
\end{tikzpicture}%
}
\caption{One explicit Chebyshev recurrent macro. The selector combines conjugate phase branches that share the real input unitary; it does not double the oracle calls. OR marks every non-success pair of signal outcomes. A controlled swap resets only those branches to the known reference. The same data register continues to the next macro. Global quadrature phase is omitted.}\label{fig:chebarchitecture}
\end{figure*}

Every other pair of signal outcomes is coherently ORed into a failure flag and swaps the data with a fresh reference register. If the success block is $B$ and failure blocks are $D_j$, unitarity gives $\sum_jD_j^\dagger D_j=I-B^\dagger B$. Discarding the environment therefore realizes
\[
B\rho B^\dagger+\Tr[(I-B^\dagger B)\rho]R
\]
without computing a square-root failure factor or the solution. Reset reuse requires $2n_r+3$ qubits and $(K-1)(n_r+3)$ ancilla resets; coherent retention requires $n_r+K(n_r+3)$ qubits. The data are never reinitialized or measured between updates. Figure~\ref{fig:chebarchitecture} depicts one complete macro update.

\begin{theorem}[Conditional separation from the original recurrence]\label{thm:separation}
Under the exact real-Julia interface above, a known gap $g$ and $q\ge1/2$, augmented-state purified distance $\eps$ is achievable with
\begin{equation}\label{eq:queryupper}
Q_{\mathrm{Ch}}=O(g^{-1/2}\log(1/\eps)).
\end{equation}
There is a family with the same interface on which the original residual channel needs $\Omega(g^{-1}\log(1/\eps))$ calls. With $g=1/(\kappa^2+1)$ this is a linear versus quadratic worst-case condition dependence for these specified recurrent channels.
\end{theorem}
\begin{proof}
Choose fixed $\eta_s\le1/2$ using \eqref{eq:chebeta}. Then $s=O(g^{-1/2})$ and Theorem~\ref{thm:cheb} gives $K=O(\log(1/\eps))$, with $2s$ queries per macro. For the lower bound, use $M_\kappa=\operatorname{diag}(1,1/\kappa)$, $b=(1,0)^{\tr}$, and start in the second solution-coordinate basis state $v$. It is orthogonal to the target and satisfies $Gv=gv$. The no-event Kraus branch after $K$ original updates has weight $(1-g)^K$ and remains orthogonal. Other branches contribute nonnegative error, so $P(\rho_K,\Pi)^2\ge(1-g)^K$. Requiring this to be at most $\eps^2$ gives $K\ge\log(1/\eps^2)/[-\log(1-g)]$, proving the separation.
\end{proof}
The simple diagonal family makes the channel lower bound transparent; it is classically trivial. The lower bound is specific to the defined original residual channel on that family, and the upper bound assumes an exact, reusable real-Julia oracle with known gap and phase operations. These are not matched dense-loading assumptions for other QLSP algorithms. In particular, the theorem neither lower-bounds QSVT, HHL, or direct classical inversion nor improves the best-known QLSP complexity. It concerns augmented-state preparation. Solution-sector extraction and full-vector classical readout can remove its benefit. For example, using the coherence residual bound to request tolerance $\tau$ can require augmented error $O(\gamma\tau)$, adding $\log(1/(\gamma\tau))$ rather than $\log(1/\tau)$ to the state-preparation estimate.

\subsection{Finite phases, tracking and noise}
The ideal achievability result does not imply exact finite-precision kernel preservation. Our numerical phase solver synthesizes $a r_s(G)$ with $a=1-10^{-12}$ and reports its floating-point Chebyshev coefficient residual. The implemented unitary still defines a CPTP channel, but the intended pure target generally leaks toward $R$. For an exact scalar attenuation,
\begin{equation}\label{eq:attenuation}
u_{k+1}\le a^2[1-q(1-\eta_s^2)]u_k+(1-q)(1-a^2).
\end{equation}
This follows by separating the target and complementary populations in $\Tr(B^2\rho)$. The resulting floor is retained in the code and must not be described as an exact stationary solution.

More generally, let implemented success block $\widetilde B$ obey $\norm{\widetilde B-B_s}\le\delta_B$, both blocks being contractions. The no-event difference has trace norm at most $2\delta_B$ and the reset-functional difference at most $2\delta_B$. Hence
\begin{equation}\label{eq:chebimplementation}
\tfrac12\norm{\widetilde\E-\E_s^{\mathrm{Ch}}}_\diamond\le2\delta_B.
\end{equation}
Over $K$ updates the trace-distance error is at most $2K\delta_B$ and purified-distance error at most $2\sqrt{K\delta_B}$, capped by one. Phase and input-unitary errors telescope: $\delta_B$ is bounded by attenuation plus polynomial-approximation error, $2s$ times the input-unitary error and the sum of phase-rounding errors. Numerical coefficient residuals are estimates, not interval-certified bounds. Maintaining an error allowance $\eps$ requires an implementation budget of order $\eps^2/K$ in this conservative inequality. This worst-case bound must not be interpreted as a demonstrated tolerance to accumulated native-gate errors; the corresponding gate-noise tests are finite, small-circuit experiments. Classical phase-synthesis cost, finite gate precision and compiled oracle realization are excluded from ideal query counts and reported separately.

For a moving target the proof of \eqref{eq:tracking} applies with $c_s=\sqrt{1-q(1-\eta_s^2)}$, because it uses target overlap rather than positivity of the no-event operator. A known uniform gap permits one filter specification across a trajectory; adaptive specifications require classical control and compilation while preserving the resident quantum state. Logical depolarization obeys \eqref{eq:noise} with the new rate. A query-coupled example uses $p_s=1-(1-p_0)^{2s}$ and remains a logical model, distinct from noise inserted into native gates.

\section{Tracking, Implementation Perturbations, and Coherence Readout}
The central recurrence is a resident density operator; the following guarantees therefore concern its tracking of a moving stationary projector, rather than convergence of a classical inverse matrix.
\subsection{Moving target under changing inputs}
Let channel $\E_{m,k}$ have target $\Pi_k$, rate $\theta_{m,k}$ and $c_k=\sqrt{1-\theta_{m,k}}$. Define purified-distance error $e_k=P(\rho_k,\Pi_k)$ and target movement $v_k=P(\Pi_k,\Pi_{k+1})$. If the next update uses channel $k+1$, then
\begin{equation}\label{eq:tracking}
e_{k+1}\le c_{k+1}\min\{1,e_k+v_k\}.
\end{equation}
The triangle inequality bounds the input error relative to the new target; Theorem~\ref{thm:accelerated} contracts its square. No reset of the data at matrix changes is assumed. If implementation changes an ideal output by purified distance at most $\xi_k$, add $\xi_k$ to the right-hand side. For uniform $c<1$, $v_k\le v$ and $\xi_k\le\xi$,
\begin{equation}\label{eq:uniformtrack}
e_K\le\min\left\{1,c^Ke_0+\frac{(cv+\xi)(1-c^K)}{1-c}\right\}.
\end{equation}
This is an upper bound; a large value does not itself prove failure. A final hold sets $v=0$ and permits convergence if implementation error is absent.

For normalized matrices with singular ratios at least $\gamma_{\min}>0$, resolvent perturbation gives
\begin{align}\label{eq:inputdrift}
\norm{\Delta z}_2\le{}&\frac{|\Delta\gamma|}{\gamma_{\min}}+
\frac{\gamma_{\mathrm{old}}}{\gamma_{\min}}\norm{\Delta\beta}_2\nonumber\\
&+\frac{\gamma_{\mathrm{old}}}{\gamma_{\min}^2}\norm{\Delta A}_2.
\end{align}
Normalization of $(z,1)$ yields the conservative bound $v\le\min(1,2\norm{\Delta z})$. The drift budget includes singular-value changes, changing conditioning and right-hand sides. Close approaches to singularity amplify its denominators; dimension alone is not the limiting parameter.

\subsection{A quantitative resident-state certificate for a moving fixed point}
\label{sec:residentcertificate}
The purified-distance tracking relation in \eqref{eq:tracking} can be
strengthened to an \emph{infidelity} certificate that explicitly charges
the Chebyshev degree, number of macro updates, moving target and
implementation noise. This also clarifies when retaining a quantum state
rather than restarting it is potentially useful.

\begin{theorem}[Time-varying signed-filter tracking with implementation error]
\label{thm:residentcertificate}
At snapshot $t$, let $\Pi_t=\ket{z_t}\bra{z_t}$ be rank one and
$R=\ket r\bra r$. Assume the \emph{ideal} trace-preserving reset map
$\E_t(\rho)=B_t\rho B_t+\Tr[(I-B_t^2)\rho]R$, where $B_t$ is a Hermitian
contraction, $B_t\Pi_t=\Pi_t$, and
$\norm{B_t(I-\Pi_t)}_\infty\le s_t<1$. Let
$q_t=\Tr(\Pi_t R)>0$,
$a_t=1-q_t(1-s_t^2)$, and
$\Delta_t=\norm{\Pi_t-\Pi_{t-1}}_\infty$.
For the signed Chebyshev response with uniform spectral gap $g_t>0$,
\begin{equation}\label{eq:degreecertificate}
s_t\le\operatorname{sech}\!\left(2d_t
                 \operatorname{artanh}\sqrt{g_t}\right).
\end{equation}
Suppose each actually implemented macro step is CPTP and deviates from
$\E_t$ in half trace norm by at most $\tau_t$, uniformly over input
states. After $K_t\ge1$ macro updates from the previous resident state,
its target infidelity $u_t=1-\Tr(\Pi_t\rho_t)$ satisfies
\begin{align}\label{eq:residentcertificate}
u_t\le \min\Bigg\{1,\;&a_t^{K_t}\min\{1,u_{t-1}+\Delta_t\}\nonumber\\
&+\tau_t\frac{1-a_t^{K_t}}{1-a_t}\Bigg\}.
\end{align}
For a freshly initialized $R$ under the same map and implementation
allowance, replace the prior factor by $1-q_t$. When the
previous \emph{certified} upper bound $\bar u_{t-1}$ satisfies
$\bar u_{t-1}+\Delta_t<1-q_t$, the resident upper certificate is
strictly tighter than the fresh upper certificate before any
saturation at one. This is a sufficient \emph{bound comparison}, not
a claim that every actual resident outcome beats every fresh outcome.
\end{theorem}
\begin{proof}
Writing $L_t=I-B_t^2$, exact kernel preservation yields
$\Tr(L_t\Pi_t)=0$ and
$L_t\succeq(1-s_t^2)(I-\Pi_t)$. Since the reset returns a fraction
$q_t$ of each failure event to $\Pi_t$,
$1-\Tr[\Pi_t\E_t(\rho)]
=u-q_t\Tr(L_t\rho)\le a_tu$, where
$u=1-\Tr(\Pi_t\rho)$. For any density operator,
$\left|\Tr[(\Pi_t-\Pi_{t-1})\rho]\right|\le\Delta_t$,
so the pre-update infidelity is at most
$\min\{1,u_{t-1}+\Delta_t\}$. The assumed half-trace-norm
implementation error changes a projector probability by at most
$\tau_t$. Iterating $K_t$ times gives \eqref{eq:residentcertificate}.
The fresh initial state has infidelity $1-q_t$.
The Chebyshev bound follows from its normalized sup-norm on
$[g_t,1]$; it does not imply efficient synthesis of $B_t$.
\end{proof}

For the illustrative post-macro depolarizing channel with per-oracle
error parameter $p_0$, $2d_t$ charged calls, and data dimension $D$,
one valid allowance is
\begin{equation}\label{eq:macroerrallowance}
\tau_t=\left[1-(1-p_0)^{2d_t}\right](1-1/D).
\end{equation}
A finite-synthesis bound can be added to this allowance by the triangle
inequality only when it is genuinely certified, not inferred from a
floating-point coefficient residual. Measuring intermediate snapshots
destroys the state in the ordinary readout model; independent copies
or trajectory replays must therefore be charged for each snapshot.
The theorem does not avoid dense spectral compilation, physical
readout costs, or comparison with leading one-shot solvers.

\subsection{Logical depolarization}
For the model $\mathcal D_p(\rho)=(1-p)\rho+pI/D$ after each ideal macro update, the exact target-overlap identity and contraction give
\begin{equation}\label{eq:noise}
e_{k+1}^2\le(1-p)(1-\theta_m)\min\{1,(e_k+v_k)^2\}+p(1-1/D).
\end{equation}
For a static target, let $\lambda=(1-p)(1-\theta_m)$. Iteration bounds the squared-error floor by
\begin{equation}\label{eq:floor}
\frac{p(1-1/D)}{p+(1-p)\theta_m}.
\end{equation}
This is a bound for a specific logical channel. It is not a physical two-qubit error threshold. We compare both fixed $p$ per macro and the illustrative charge-coupled model $p_m=1-(1-p_{\mathrm{base}})^\ell$. The latter accounts for rising exposure phenomenologically; it does not reproduce noise interleaved with every physical gate.

\subsection{Implementation perturbations}
For valid contractions with the same reference, telescoping powers gives
\begin{equation}\label{eq:lip}
\norm{q_m(G)-q_m(\widetilde G)}\le\ell\norm{G-\widetilde G}.
\end{equation}
Let the left-side norm be $\delta$. Positive square-root continuity implies $\norm{C_m-\widetilde C_m}\le\sqrt\delta$, without a false commutativity assumption. For any joint state, the no-event term differs in trace norm by at most $2\sqrt\delta$ and the reset functional by at most $\delta$. Thus
\begin{equation}\label{eq:diamond}
\tfrac12\norm{\E_m-\widetilde\E_m}_{\diamond}\le\zeta:=\min\{1,\sqrt\delta+\delta/2\}.
\end{equation}
The output purified distance is at most $\sqrt{2\zeta}$, usable as $\xi$ in \eqref{eq:uniformtrack}. A directly compiled dilation with operator-norm error at most $\eta$ instead has output purified-distance error at most $\eta$ for a common environment input. Product-unitary errors sum by telescoping, including controlled calls and swaps. Larger degree can therefore amplify implementation sensitivity while reducing macro updates.

\subsection{Coherence readout and absolute scale}
For real solutions use the reference coherences:
\begin{equation}\label{eq:readout}
\widehat x_i=\frac{\norm b_2}{\gamma\norm M_2}
\frac{\operatorname{Re}\rho_{ir}}{\rho_{rr}}.
\end{equation}
At the fixed state, $\rho_{ir}/\rho_{rr}=z_i$, so sign and magnitude follow without fitting a scale to the classical solution. If $e=P(\rho,\Pi)<1/2$, trace-distance bounds imply $|\rho_{rr}-q|\le e$ and $\norm{\rho_{sr}-qz}\le e$. Consequently
\begin{equation}\label{eq:resread}
\frac{\norm{M\widehat x-b}_2}{\norm b_2}\le\frac{2e}{\gamma(1/2-e)}.
\end{equation}
This explains why a favorable augmented fidelity alone can be inadequate for inverse-action precision at large $\kappa$.

Readout requires measurements. With absolute error at most $t$ on each of $N$ real coherences and the reference probability, $e+t<1/2$ gives the conservative bound
\begin{equation}\label{eq:shots}
\frac{\norm{M\widehat x-b}}{\norm b}\le
\frac{2e+(\sqrt N+1)t}{\gamma(1/2-e-t)}.
\end{equation}
Coherences are measured by observables $(\ket i\bra r+\ket r\bra i)/2$, with outcomes in $[-1/2,1/2]$. Hoeffding's inequality and a union bound require at least $\log[2(N+1)/\beta_f]/(2t^2)$ independent samples per setting for failure probability $\beta_f$. Each sample requires another state preparation and recurrence execution. We do not infer full-vector readout efficiency from exact simulator coherences.

\section{Whole-Algorithm Cost and Readout Confidence}\label{sec:completecost}
An implementation advantage is meaningful only under common accuracy, confidence, input and hardware assumptions. We report raw queries, CX gates, depth, width and reset requirements separately from a time model. Define
\begin{equation}\label{eq:totalcost}
\begin{aligned}
T_j^{\mathrm{total}}={}&\frac{T_j^{\mathrm{pre}}+T_j^{\mathrm{phase}}+T_j^{\mathrm{compile}}}{R_c}\\
&+S_j\bigl[T_j^{\mathrm{prepare}}+T_j^{\mathrm{circuit}}+T_j^{\mathrm{readout}}\bigr].
\end{aligned}
\end{equation}
Here $R_c$ counts reported solves of the same compiled $M,b$ instance, while $S_j$ counts circuit executions per reported solve. The update count is selected offline from scored simulator trajectories, not by a demonstrated hardware stopping rule. Diagnostic spectral propagation and reference scoring are recorded separately and excluded from compiler time; actual full-circuit assembly is included. Changing $b$ generally changes this augmented residual. Reuse does not make arbitrary changing inputs free. Classical times are measured locally; gate durations are declared assumptions scheduled with their qubit dependencies. The model includes actual synthesized coherence-measurement basis rotations and allows parallel resets on independent qubits.

Known reference-preparation gates are included in the native circuit. The reported timing examples assume zero-state qubits are supplied at the start of an execution and assign no additional preparation latency. Platform initialization between shots, queueing and classical control latency are not measured. Such overheads must be added before interpreting the model as device time-to-solution.

For residual confidence target $\tau$, solve \eqref{eq:shots} for the observable precision:
\begin{equation}\label{eq:precisionbudget}
t\le\frac{\tau\gamma(1/2-e)-2e}{\sqrt N+1+\tau\gamma}.
\end{equation}
If the numerator is nonpositive, the bound cannot certify readout and no finite shot count is assigned. Otherwise each of the $N+1$ settings is charged the Hoeffding budget. The matched small cost study requires both the joint $10^{-6}$ simulator accuracy criterion and $e\le10^{-2}\gamma/200$, reserving state-error budget for a common $\tau=10^{-2}$ readout certificate. Tighter $10^{-6}$ readout failures are reported. These conservative shot bounds can be very large and are not executed hardware shots.

For the illustrative $K_0=99$, $K_{16}=13$ and $\ell=17$ charge 99 versus 221 primitive calls for the controlled-baseline construction. If both methods had the same bulk management latency $t_R$ per macro and extra polynomial control latency $h$, an advantage would require
\begin{equation}\label{eq:break_even}
t_R>\frac{122}{86}t_U+\frac{13}{86}h.
\end{equation}
This conditional threshold is not a measured reset advantage: actual ancilla counts and reset topology differ. The resource scheduler uses those actual circuits instead. Similarly, the earlier dense-compiler ratio $3366/442\approx7.62$ is an instance- and compiler-specific CX saving, not an oracle-query or end-to-end hardware speedup.


\section{Implementation and Reproducible Experimental Design}
\subsection{Source, simulator and validation levels}
The implementation and the archived numerical evidence are in the public GitHub repository~\cite{r35}. Python 3.12, NumPy 2.3.5, SciPy 1.17.0, Qiskit 2.6.0 and Aer 0.17.2 are pinned \cite{r36}; plots use Matplotlib 3.10.8. Seeds, exact commands, inputs, source snapshot and checksums are archived. A fresh full campaign is identified in the generated evidence table rather than inferred from an earlier run. Software validation, accuracy outcomes and hardware evidence are separate labels.

Large static cases, including all order-1000 examples, use exact renewal/FFT observables of the reduced channel in its residual spectral basis. They are classical simulations of quantum maps, not gate-level executions of a 1000-dimensional quantum circuit. Dynamic cases preserve the physical density matrix through exact changes of simulation basis. The classical reference solution is used only to score observations, never to prepare the recurrence or correct its iterate. Conditional solution fidelity is evaluated on the full mixed solution-sector density, not a normalized vector of readout coherences.

Small independent checks construct complete circuits, evolve the full data/environment state and trace out the environment. These unitary and density-matrix simulator checks assess the mathematical implementation, not physical-device accuracy. A distinct deferred-reset study composes controlled original-unitary calls. Native counts use actual transpilation in an all-to-all native basis without device calibration. Large circuits are rejected by explicit budgets rather than silently replaced by a reduced-channel simulation while retaining a native label.

\subsection{Static design and metrics}
The matched grid uses $N\in\{2,4,10,100,1000\}$, $\kappa\in\{2,10,100,1000\}$, degrees $m\in\{0,1,2,4,8,16\}$, and symmetric positive-definite, nonsymmetric, symmetric-indefinite and clustered-spectrum matrices. Spectra are controlled, not assumed from random matrix entries. Initial states include the reference, a basis state, a seeded coherent random state and a mixed state on the applicable grids. Baseline and polynomial methods use identical input bytes, thresholds and initial states.

The joint accuracy criterion is relative residual at most $10^{-6}$ and conditional solution infidelity at most $10^{-6}$. Static budgets are 512, 4096, 16384 and 16384 macro updates for the four increasing condition numbers. Recorded trajectories include the initial residual, final residual, purified distance, fidelity, solution-sector probability, $x_1$, absolute $x_1$ error and trace error. A reported convergence time is the first recorded point after which all recorded points satisfy the criterion; finite sampling can overestimate the actual first crossing and does not certify every unrecorded iterate.

Native complete convergence compilation is separately limited to 128 updates and 50000 CX gates. Resource rejections, failed criteria and timeouts are data. Preprocessing time is separated from spectral propagation, transpilation and simulation time. All times are local classical measurements; they are not quantum execution times or asymptotic benchmark estimates.

\subsection{Dynamic and noise design}
Changing-spectrum trajectories use $N=10,100,1000$, $m=0,4,16$, and logical probabilities $0,10^{-4},10^{-3}$. The condition sequence is $2,10,100,1000,1000,100,10,2$. Each of the first seven snapshots lasts 32 updates; the final matrix is held for 512, giving 736 updates and a hold beginning at 224. Singular values, scale, orientation and the right-hand side change, with an abrupt central jump. Target and measured $x_1$, full-vector residual and tracking error are plotted for every case.

Separate high-conditioning tests use $\kappa=100,1000$, the three large sizes, degrees $0,4,16$, two base probabilities and both noise conventions. Native gate-noise tests use $N=2$, $\kappa=2,100$, 16 macro updates, CX depolarization $0,10^{-4},10^{-3}$, and a post-reset bit-flip probability $10^{-3}$ in nonideal cases. Other gates and measurement are ideal. No physical backend calibration is used. The original rotating benchmark is retained in the full archive.

\subsection{Matched dissipative solver and equal-cost audit}
The Shang comparison uses the same normalized matrices, right-hand sides, initial reference state, output readout, sampled times and accuracy criteria. It reports both the original continuous generator and a rate-normalized convention. Since the unnormalized augmented residual has squared norm at most two, the rate-normalized generator is $\mathcal L/2$; a normalized time $t$ corresponds to original-generator time $t/2$. Continuous time is not a native gate count.

The discrete and continuous solvers return the same classical inverse action through separately calibrated coherence readouts, but their augmented stationary states differ. Shang uses reference weight one in $[A,-\beta]$; our recurrence uses $\gamma$ in $[A,-\gamma\beta]$. Thus Shang's stationary solution-sector probability is $\norm{A^{-1}\beta}^2/(1+\norm{A^{-1}\beta}^2)\ge1/2$, whereas ours is at most $1/2$. This encoding tradeoff is reported rather than hidden by normalization. The comparison examines physical solution residuals, not overlaps between different augmented targets. Table~\ref{tab:comparison} contrasts the encodings and recurrence models. The separate equal-call audit fixes 16384 primitive calls and allocates $\lfloor16384/(m+1)\rfloor$ macros. Extra controlled-call synthesis adds cost beyond this accounting.

\subsection{Accuracy-matched, held-out total-cost protocol}
\label{sec:costprotocol}
To distinguish tracking feasibility from resource advantage, we evaluate an independent $360$-scenario factorial grid at $N=2$, $\kappa=10$, a 12-snapshot horizon, and at most 256 primitive calls per input snapshot. The three families change right singular vectors, the right-hand side alone, or introduce abrupt rotations. Angular drift is $0,0.01,0.05,0.2,0.6$ radians per snapshot; logical per-oracle noise is $0,10^{-6},10^{-5},2\times10^{-5},5\times10^{-5},10^{-4}$; destructive readout occurs at the final snapshot only or every 1, 3, or 6 snapshots. These input controls are distinct from the measured operator-norm drift between target projectors.

Each method selects its polynomial degree and recurrent update count using only validation seeds 23 and 29, under the same per-snapshot query ceiling. The frozen policy is evaluated on disjoint held-out seeds 41 and 43. A readout checkpoint passes if the conditional solution infidelity is at most $0.01$ and relative residual at most $0.05$; a policy qualifies when at least 90\% of held-out checkpoints pass. These deliberately exploratory criteria \emph{do not} replace the stricter $10^{-3}$ target of Table~\ref{tab:temporalpaired}. One-shot filtering is postselected, with expected retries charged separately.

For $H=16$ assumed shots in each of $N+1$ coherence settings, the cost model charges $S=(N+1)H=48$ independently prepared copies per readout checkpoint. It does \emph{not} simulate shot outcomes: acceptance is scored against exact small density matrices. Compilation is optimistically amortized over distinct input snapshots. Classical reference solutions score the outputs but never drive channel updates.

\section{Results: Convergence, Resources, and Failure Modes}

\subsection{Static convergence and conditioning}
Across the 1416 matched static trajectories, 770 meet the joint sampled accuracy criterion and 646 fail it within budget. Tables~\ref{tab:condition} and \ref{tab:degree} aggregate the actual records. All condition-2 and condition-10 cases pass. At condition 100, only degree 16 passes across its tested initial states and families. At condition 1000, no degree passes within 16384 macro updates. These are finite-budget results, not exceptions to the asymptotic attraction theorem.

\begin{table}[!tbp]\centering\caption{Static outcomes by condition number. Accuracy failure is retained independently of software validation.}\label{tab:condition}
\begin{tabular}{lrrr}\toprule
Condition & Cases & Pass & Fail\\\midrule
2 & 372 & 372 & 0\\
10 & 336 & 336 & 0\\
100 & 372 & 62 & 310\\
1000 & 336 & 0 & 336\\
\bottomrule\end{tabular}

\end{table}
\begin{table}[!tbp]\centering\caption{Static outcomes by polynomial degree. Median first sampled update is conditional on a passing trajectory; the successful population changes with degree.}\label{tab:degree}
\begin{tabular}{lrrr}\toprule
Degree & Pass & Cases & Median first update\\\midrule
0 & 118 & 236 & 112\\
1 & 118 & 236 & 60\\
2 & 118 & 236 & 41\\
4 & 118 & 236 & 25\\
8 & 118 & 236 & 22\\
16 & 180 & 236 & 131\\
\bottomrule\end{tabular}

\end{table}

\begin{figure*}[!tbp]\centering\includegraphics[width=.96\textwidth]{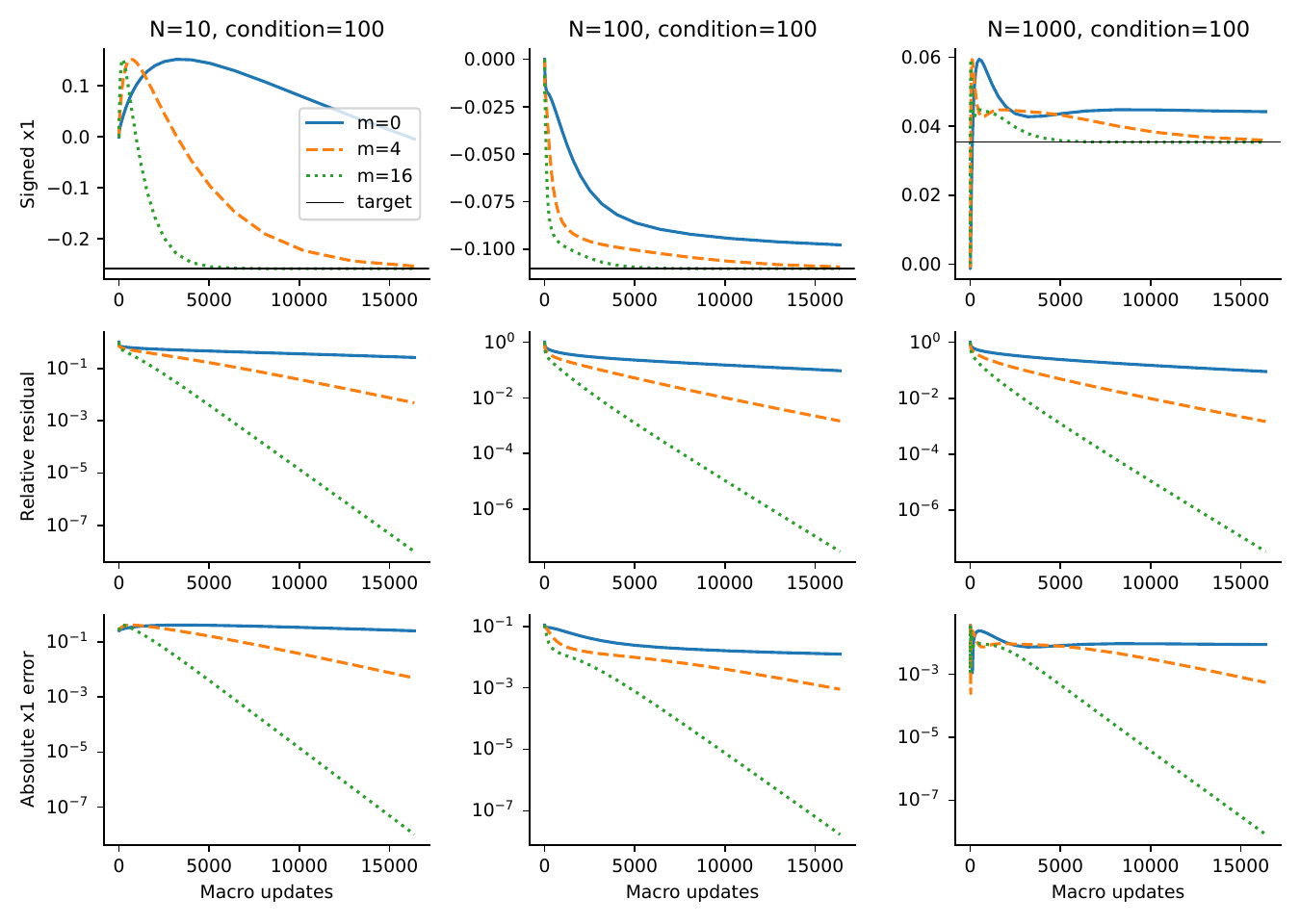}\caption{Matched nonsymmetric matrices at condition 100. The horizontal target for $x_1$ is the classical reference used only for scoring. Baseline and accelerated methods start from the same reference state and use identical matrices. Each column is a different dimension. Curves are actual sampled records.}\label{fig:static}\end{figure*}

Figure~\ref{fig:static} shows the actual signed first component, relative residual and absolute component error. This prevents apparent convergence of one component from standing in for a full-vector accuracy test. The complete data contain every initial state and matrix family, together with undefined initial readouts where appropriate. A larger degree can accelerate residual relaxation in macro time, but the difficult tail of the spectrum still controls finite-budget behavior.

The maximum independent two-update complete-circuit density discrepancy is $8.771\times10^{-15}$ across 72 comparisons. This validates the reduced-channel representation for those small built circuits. It does not make the 1000-dimensional renewal simulation a 1000-dimensional native-circuit execution. Native construction yields 72 compiled cases and 48 explicit large-case resource rejections.

\subsection{Moving-target tracking and final convergence}
\begin{figure*}[!tbp]\centering\includegraphics[width=.96\textwidth]{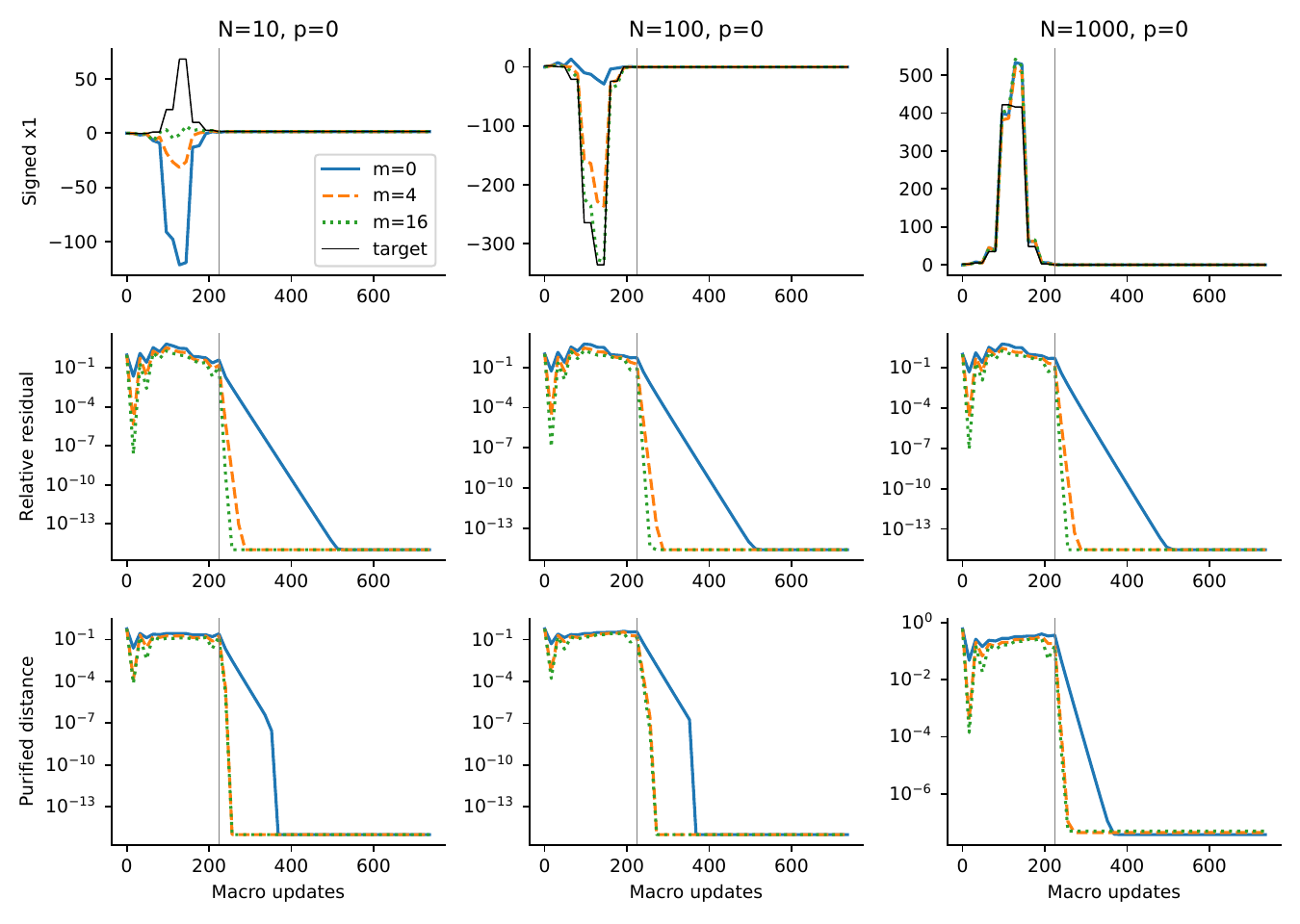}\caption{Ideal changing-spectrum trajectories with changing right-hand sides and abrupt jumps. The final hold starts at update 224 (vertical line). Top: actual and target $x_1$. Middle: complete-vector relative residual. Bottom: augmented-state purified distance. These are reduced-channel simulations at all three dimensions.}\label{fig:dynamicideal}\end{figure*}

\begin{figure*}[!tbp]\centering\includegraphics[width=.96\textwidth]{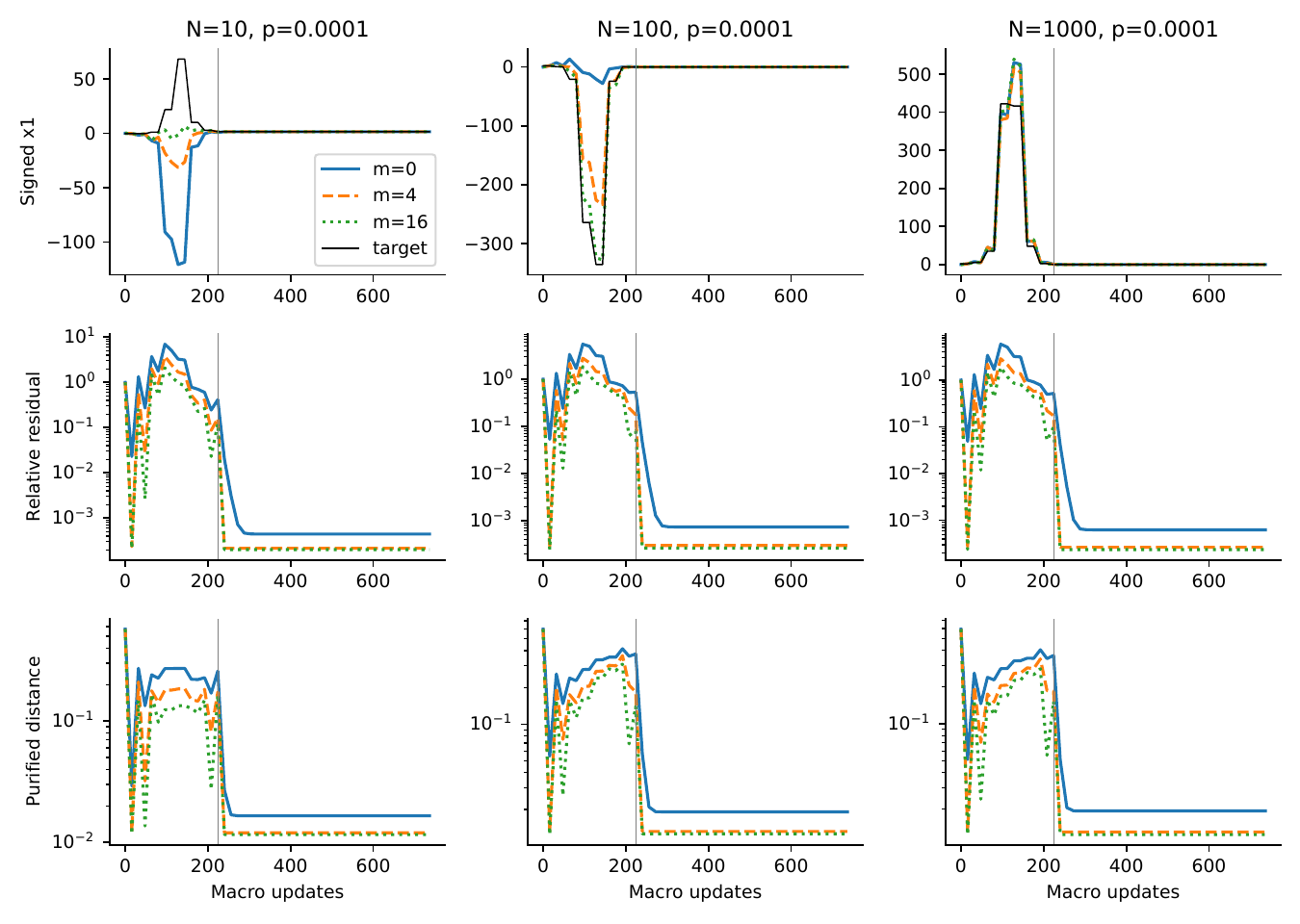}\caption{The same matrix trajectories with logical depolarization $p=10^{-4}$ after every macro update. The first-component plot alone can hide a persistent full-vector residual. Logical noise remains distinct from physical gate-level calibration. Individual plots for every probability, dimension and degree are included.}\label{fig:dynamicnoise}\end{figure*}

Figures~\ref{fig:dynamicideal} and \ref{fig:dynamicnoise} compare ideal and logically noisy trajectories. The 27 changing-spectrum trajectories include nine ideal and eighteen logically noisy cases. All nine ideal cases recover after the final well-conditioned hold. The recorded hold-to-convergence delays are 112, 32 and 16 updates for degrees 0, 4 and 16, respectively, in each tested dimension. Sampling is every 16 updates, so these are sampled delays. The matrix is not held at condition 1000 when this recovery occurs; its final condition is two.

The transient jumps are substantial: the measured error can increase when the matrix and right-hand side change, even though the density state is retained. Final noise floors prevent the strict $10^{-6}$ accuracy criterion. Thus ideal final-hold recovery and robust noisy tracking are different experimental outcomes. The archive retains the original rotating-matrix study and the new changing-spectrum, changing-condition study, rather than replacing the former with a more favorable trajectory.

Separate complete-circuit checks use two explicit nonsymmetric $2\times2$ matrices and right-hand sides, four updates, degrees 0, 4 and 16, reference and complex coherent initial states, and both coherent environment retention and reset reuse. Twelve circuits are compared at all five prefixes, including the initial state. The generated audit reports their actual density discrepancies and native resources. This check verifies changing-gate consistency over short schedules; it does not establish long-time convergence under arbitrary changes.

\subsection{Paired temporal recurrence versus fresh and one-shot filtering}
\label{sec:matchedtemporal}
We separately evaluate the specific value of persisting the physical data
state through a changing matrix and right-hand side. For each snapshot $t$,
let $B_{d,t}=r_d(G_t)$, $R=\ket r\bra r$, and $\E_{d,t}$ be the signed
Chebyshev reset channel in (19). A \emph{resident} method applies
$\rho_t^{\rm res}=\E_{d,t}^{K}(\rho_{t-1}^{\rm res})$ without
reinitializing the data register, whereas a \emph{fresh} baseline applies
the identical map from $R$ at every snapshot,
$\rho_t^{\rm fresh}=\E_{d,t}^{K}(R)$.
Both charge $2dK$ primitive calls and share the same per-snapshot
ceiling $Q$, with $K=\lfloor Q/(2d)\rfloor$.
The third comparator applies $B_{s,t}$ once to $R$ and postselects the
success branch, where $s=\lfloor Q/2\rfloor$; the success probability
$p_t=\Tr(B_{s,t} R B_{s,t}^{\dagger})$ entails
an expected $2s/p_t$ calls per accepted output, so this is \emph{not}
an equal-expected-cost unconditional channel comparison.

The controlled study uses $N\in\{2,4\}$, condition numbers $2$ and $10$,
query ceilings $Q\in\{32,64,128\}$, two seeded realizations,
six snapshots, gradual versus abrupt rotations of the right singular
vectors, changing right-hand sides, and logical per-oracle depolarization
$p\in\{0,10^{-3}\}$. We require both relative residual and conditional
solution infidelity at most $10^{-3}$. Classical reference solves are used
only for scoring, never for state updates. Every case, including a tied or
failed one, contributes to Table~\ref{tab:temporalpaired}.

\begin{table}[!tbp]\centering\small
\caption{Matched temporal reduced-channel experiment: numbers of snapshots
passing both accuracy criteria. Each method is evaluated at all 576
snapshots. One-shot conditioning requires a separate success/retry budget.
These are classical simulations of ideal or logically noisy quantum maps,
not native-gate or QPU experiments.}
\label{tab:temporalpaired}
\begin{tabular}{lrrr}\toprule
Case subset & Resident & Restarted & One-shot\\\midrule
All snapshots (576) & 154 & 87 & 194\\
Gradual change (288) & 93 & 42 & 96\\
Abrupt change (288) & 61 & 45 & 98\\
Logical $p=10^{-3}$ (288) & 0 & 0 & 0\\\bottomrule
\end{tabular}
\end{table}

Among the 96 paired configurations, resident operation wins on the
number of accuracy-passing snapshots in 18, ties in 78, and loses in none.
Direct one-shot filtering passes more snapshots overall, but its median
expected accepted-state query multiplier is $1.54$ (range $1.01$--$1.99$)
relative to one single attempt at the reported ceiling. In particular,
all three methods fail the strict criterion under the tested nonzero logical
noise, so the data do \emph{not} establish robust noisy tracking.
The resident advantage is conditional on this benchmark, access model,
threshold, and finite simulation grid; no universal superiority or
hardware-latency advantage follows.

For independently tested real-Julia QSP signal blocks at $\kappa=2$ and
$10$, the maximum success-block discrepancies are respectively
$1.34\times10^{-14}$ and $2.82\times10^{-14}$. Their empirical coefficient
residuals are about $1.82\times10^{-14}$ and $1.79\times10^{-14}$,
not certified interval bounds. Simulated query-coupled depolarization
adds infidelity biases of approximately $0.00895$ and $0.02997$ in those
small checks. These are logical-noise results, distinct from native-gate
noise and real hardware calibration. The separate archived outputs retain
per-snapshot residuals, conditional infidelities, acceptance probabilities,
trace checks, and the independently reproduced filter-polynomial checks.

\subsection{Expanded temporal and native-gate stress tests}
\label{sec:expandedtracking}
An additional, fully retained factorial experiment,
\texttt{qcenn\_mi.extended\_tracking\_benchmark}, separately
tests the predictions of Theorem~\ref{thm:residentcertificate}
without changing the original 96-case study. It evaluates five
matrix/RHS families (rotating, RHS-only, singular-spectrum drift,
rank-two shocks and periodic drift), $N\in\{2,4\}$,
$\kappa\in\{2,10\}$, budgets $Q\in\{32,64,128\}$,
six and twelve snapshots, gradual and abrupt changes, three seeds,
and logical per-oracle noise rates
$p_0\in\{0,10^{-5},10^{-4},10^{-3}\}$.
The resulting 2880 paired configurations contain 25\,920
snapshots \emph{per method}. The source-pinned full-evidence
archive and SHA-256 manifest are available from
\url{https://github.com/vtavakkoli/QCeNN-MI/actions/runs/40408826555}.
All evaluated reduced-channel target infidelities satisfied the
new \emph{upper certificates}; this validates the bound on these
fixtures, not its novelty relative to every prior theorem.

\begin{table}[!tbp]\centering\small
\caption{Expanded time-varying reduced-channel benchmark,
25\,920 snapshots per method, $2880$ paired configurations.
Passing both accuracy criteria is counted; direct filtering is
postselected and has a separate accepted-state cost.
No native-device advantage is claimed.}
\label{tab:expandedtemporal}
\begin{tabular}{lrrr}\toprule
Method & Passing & Failing & Pass rate\\\midrule
Resident Chebyshev & 7254 & 18\,666 & 28.0\%\\
Fresh Chebyshev & 3653 & 22\,267 & 14.1\%\\
One-shot Chebyshev & 7530 & 18\,390 & 29.1\%\\\bottomrule
\end{tabular}
\end{table}
The resident method improves the number of passing snapshots
over fresh recurrence in $608$ configurations, ties in $2271$,
and loses in $1$. The pooled paired mean difference is
$1.250$ passing snapshots per configuration; a trajectory-level
percentile bootstrap reports $[1.147,1.355]$.
This interval is \emph{descriptive for the fixed synthetic grid},
because cases share fixture families and seeds and are not an
independent random sample of deployment conditions. All
$720$ cases at $p_0=10^{-4}$ and all $720$ at $p_0=10^{-3}$
have \emph{zero paired pass-count difference}; there is no
demonstrated robust resident advantage in these noisy slices.
Although direct filtering has the largest raw pass count,
retries and readout make the per-accepted-copy work different.
Measuring a moving resident output consumes the data state in
the ordinary measurement model; obtaining every snapshot
requires independent copies or replay of the input schedule,
which can nullify nominal state-preparation reuse.

A separate native-gate audit
(\path{qcenn_mi.native_limitation_audit}) compiles the
small complete real-Julia QSP reset circuit under line
connectivity and applies explicitly \emph{synthetic}
gate-local depolarization to selected cases.
It measures CX/depth pairs $(380,867)$ and $(1238,3067)$
for $N=2$ at $\kappa=2,10$, and $(1782,3816)$ and
$(6424,14134)$ for $N=4$ at the same conditions.
These counts describe one complete small-circuit macro,
not $N=1000$ native execution. The module separately records
reference preparation, dense SVD/phase synthesis, and the
gate-local no-error-event likelihood; readout-error exposure
is not equated to tomography-verified output fidelity.
No calibrated quantum hardware or end-to-end speedup is claimed.
The separate full source-pinned native archive is in the same
GitHub Actions workflow.

\subsection{Total-cost frontier with destructive readout}
\label{sec:costfrontier}
The factorial analysis explicitly charges the consequence of observing a resident state. Let $\mathcal{T}$ be the zero-based destructive-readout checkpoints and $S$ the independent copies required per checkpoint. With validation-selected degree $d_j$ and macro count $K_j$, the resident method must propagate a new copy through every prior input to reproduce the state at $t$, whereas a fresh method prepares only the requested snapshot:
\begin{equation}\label{eq:replayqueries}
\begin{aligned}
Q_{\mathrm{res}}&=2d_{\mathrm r}K_{\mathrm r}S\sum_{t\in\mathcal{T}}(t+1),\\
Q_{\mathrm{fresh}}&=2d_{\mathrm f}K_{\mathrm f}S|\mathcal{T}|.
\end{aligned}
\end{equation}
The full model additionally charges reference preparations, macro resets, readout executions, and compilation of distinct input snapshots, using nonnegative equivalent-unit weights $(w_q,w_R,w_P,w_M,w_C)=(1,2,5,0.1,10)$; the precise break-even inequality appears in the supplement. If the two methods use the same $d,K$ and require destructive replays, residence cannot strictly reduce the assumed cost. We compare \emph{cost only among methods that independently pass the same held-out accuracy criterion}, not between a qualified and a failed output.

\begin{table}[!tbp]\centering
\caption{Recurrent-state cost frontier: 360 synthetic scenarios, $N=2$, $\kappa=10$. Costs are hypothetical equivalent units; qualification requires at least 90\% of held-out checkpoints to meet both exploratory accuracy thresholds.}
\label{tab:costfrontier}
\setlength{\tabcolsep}{4pt}
\begin{tabular}{lr}\toprule
Held-out recurrent-method outcome & Scenarios\\\midrule
Both qualify: fresh cheaper & 184\\
Both qualify: resident cheaper & 0\\
Resident qualifies, fresh does not & 74\\
Fresh qualifies, resident does not & 4\\
Neither recurrent method qualifies & 98\\\midrule
Total & 360\\\bottomrule
\end{tabular}
\end{table}
For the 184 scenarios where both recurrent methods qualify, the median resident-to-fresh total-cost ratio is $3.12$ (range $2.19$--$8.57$). Resident tracking alone satisfies the recurrent accuracy criterion in 74 other scenarios, notably 42 of the 72 cases at input-drift control $0.2$; at $p=10^{-4}$ neither recurrent method qualifies in any of 60 scenarios. Including the direct postselected control and its expected retries, the least-cost qualified choice is direct filtering in 216 cases, fresh recurrence in 36, and resident recurrence in 20; no method qualifies in 88. The 20 resident least-cost choices arise in scenarios without a qualified fresh comparator and do not imply a resident cost advantage when both qualify.

These outcomes use two held-out seeds, exact density-state accuracy rather than sampled measurement outcomes, and an uncalibrated cost model. They demonstrate \emph{conditional tracking feasibility} but not an end-to-end or native-device total-cost benefit. Results by drift and noise, the detailed cost inequality, and source limitations are provided in the supplement.

\subsection{High-conditioning logical failures and gate noise}
\begin{figure}[!tbp]\centering\includegraphics[width=.94\columnwidth]{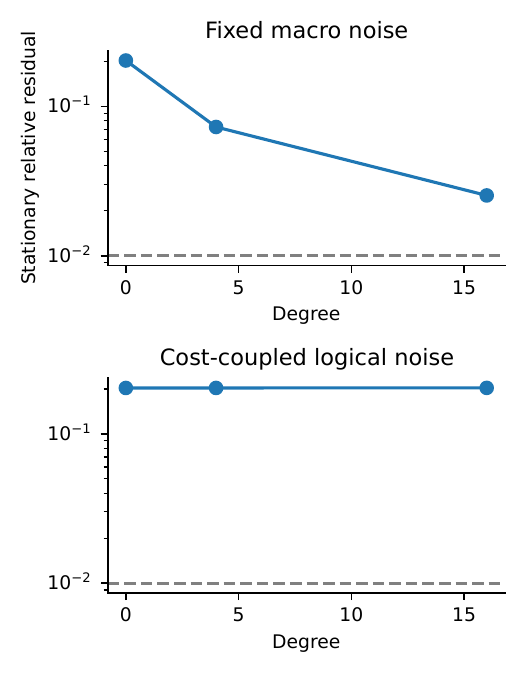}\caption{Condition-100, order-100 logical stationary residual at base probability $10^{-4}$. The dashed line is the $10^{-2}$ criterion. Fixed-macro and cost-coupled noise are different assumptions, both simulated. Neither panel reports physical gate errors.}\label{fig:noise}\end{figure}

None of the 72 high-conditioning logical-noise transient or stationary residuals reaches $10^{-2}$. In the fixed-macro model, greater degree can lower the noise floor because each update contracts more strongly while the specified noise is unchanged. Under charge-coupled noise the gain largely disappears. Figure~\ref{fig:noise} uses one matched condition-100 example to show this distinction. These failures preserve the earlier condition-100 negative evidence.

The 18 native gate-noise cases are limited to two-dimensional matrices. For the condition-2 degree-16 case after 16 updates, the ideal residual is approximately $1.029\times10^{-8}$; adding CX depolarization and reset error degrades the result. At condition 100, even the ideal 16-update circuit has substantial residual. These finite circuits do not validate the large logical-noise model physically. Exact per-case residuals, fidelities, resources and timing remain in the raw gate-noise table.

\subsection{Comparison against the established dissipative solver}
\begin{figure*}[!tbp]\centering\includegraphics[width=.92\textwidth]{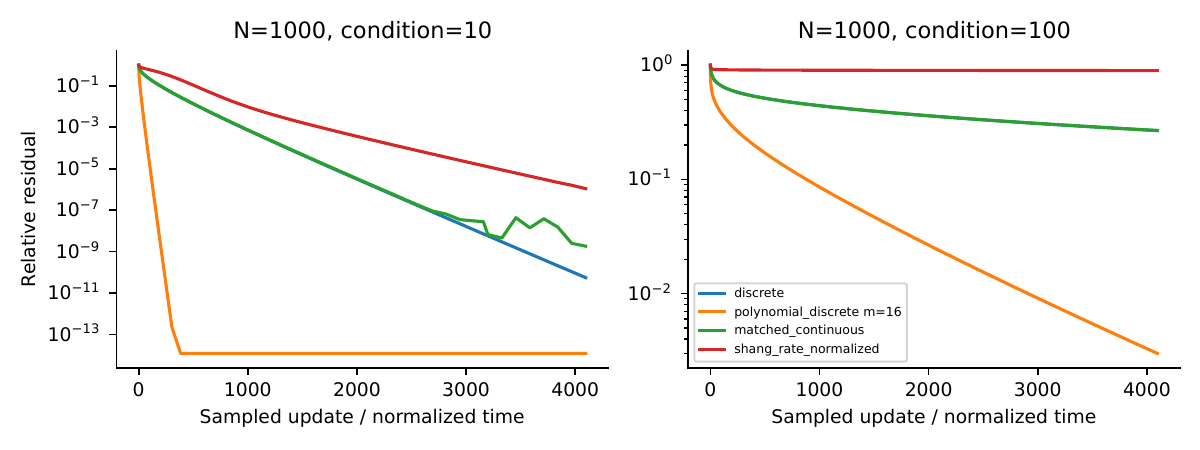}\caption{Matched comparison with a same-$F$ continuous control and Shang's separate residual encoding at rate $\mathcal L_S/2$. Discrete macro time, continuous mixing time and primitive circuit calls are distinct costs. The plot supports trajectory comparison, not an end-to-end quantum speedup benchmark.}\label{fig:prior}\end{figure*}

Figure~\ref{fig:prior} compares the baseline discrete recurrence, its degree-16 extension and continuous dissipative trajectories on identical 1000-dimensional inputs. At condition 10, the original discrete channel and the continuous control using the same residual $F$ first meet the sampled criterion at 2304, whereas the polynomial method does so at 140 macro updates. The latter uses 17 primitive calls per macro, so 140 macros charge 2380 calls before controls. Shang's separate encoding is integrated at generator $\mathcal L_S/2$ and does not cross in this window. Continuous time cannot be identified with either gate count.

At condition 100, no compared method meets the joint target in the 4096-step comparison window. The degree-16 final relative residual is about $3.00\times10^{-3}$, versus $2.68\times10^{-1}$ for the original recurrence and its same-$F$ continuous control, and $8.96\times10^{-1}$ for Shang's rate-normalized trajectory. These values must not be conflated with the longer 16384-update static experiment in which degree 16 passes. The matched numerical comparison is provided, but an equally optimized complete digital implementation of the continuous solver is unavailable here.

\begin{table*}[!tbp]\centering\caption{Conceptual comparison under matched augmented-state output conventions. Query complexities of other access models are not used as measured runtimes.}\label{tab:comparison}
\begin{tabular}{p{.18\textwidth}p{.24\textwidth}p{.24\textwidth}p{.24\textwidth}}\toprule
Property & Continuous residual reset & Original discrete recurrence & Polynomial recurrence\\\midrule
Residual & Augmented $[A,-\beta]$ & Scaled augmented $F$ & Same input $F$; response $q_m(G)$\\
Evolution & Lindblad semigroup & Collision CPTP map & Deferred-event CPTP macro map\\
Stationary output & Inverse action, reference weight one & Inverse action, reference weight $\gamma$ & Same state as original discrete method\\
Rate mechanism & Dissipative spectral gap & Positive residual gap & Gap changed to $1-(1-g_+)^{m+1}$\\
Extraction & Solution weight at least $1/2$ & Solution weight in $[1/(\kappa^2+1),1/2]$ & Weight unchanged by degree\\
Access & Block-encoding/state-preparation assumptions in prior algorithm & Dense compiler in this implementation & Dense transformed compiler or controlled $U_0$ calls\\
Temporal evidence here & Matched reduced continuous dynamics & Resident-state trajectory & Same trajectory, new polynomial response\\
Noise evidence here & Logical-channel comparator & Logical and small synthetic gate tests & Logical and small synthetic gate tests\\\bottomrule
\end{tabular}\end{table*}

\subsection{Total cost, compilation and equal-call outcomes}
\noindent\textbf{Classical-solver comparison scope.} Classical factorization is already required by the dense spectral compiler. The reported recurrence experiments do not provide a matched wall-clock time-to-solution comparison against an optimized classical LU/QR/SVD solver with the same output task, accuracy and hardware. Such a benchmark would have to charge matrix acquisition and full-vector readout, and distinguish a classical output from a quantum encoded state. Consequently no classical runtime or memory advantage is inferred from the recurrence tables.

\begin{figure}[!tbp]\centering\includegraphics[width=.94\columnwidth]{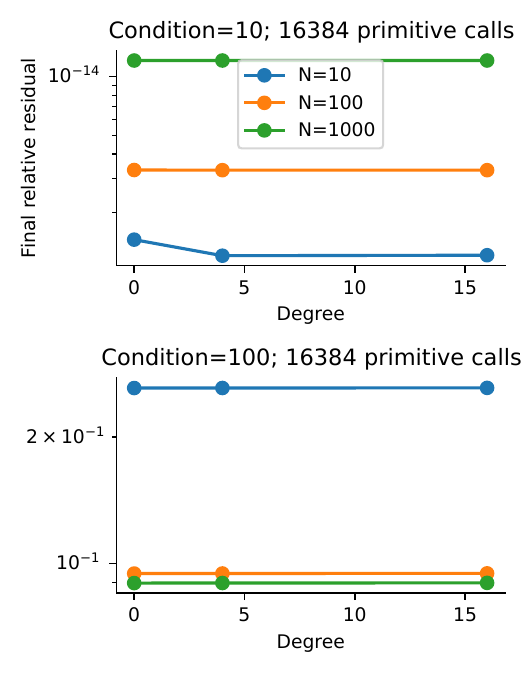}\caption{Equal primitive-call audit. Small differences at condition 100 do not establish acceleration; all cases there fail the accuracy target. At condition 10 all tested degrees pass. The audit measures the specified finite budget and does not claim optimality of either method.}\label{fig:equal}\end{figure}

The equal-call audit has nine passing condition-10 cases and nine failing condition-100 cases. With 16384 primitive calls, the condition-100 order-1000 residuals are approximately 0.08976, 0.08979 and 0.08985 at degrees 0, 4 and 16. There is no residual improvement in this matched equal-call test. Figure~\ref{fig:equal} displays all three tested sizes. Extra controls, swaps and readout would add cost beyond this call accounting.

\begin{figure}[!tbp]\centering\includegraphics[width=.94\columnwidth]{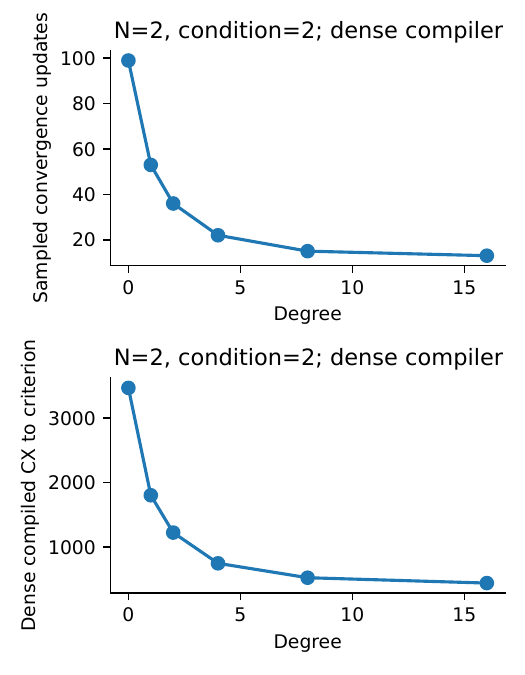}\caption{Degree versus recurrent updates and dense compiled CX count for the selected small fixture. Both are actual compiled results. This favorable dense-synthesis example coexists with the equal-primitive-call negative result.}\label{fig:resources}\end{figure}

Dense transformed compilation can have a different empirical tradeoff. In the two-dimensional condition-2 fixture, the baseline has 35 CX gates and depth 114 per update, reaching the sampled criterion after 99 updates and 3465 CX gates. Degree 16 reaches it after 13 updates and 442 CX gates in this dense compiler. Table~\ref{tab:resources} and Fig.~\ref{fig:resources} report the actual compiled cost. The classical transformed-unitary synthesis is charged separately and does not demonstrate an oracle-efficient advantage.

\begin{table}[!tbp]\centering\caption{Measured dense compiled resources for the same two-dimensional condition-2 fixture. Complete-convergence circuits satisfy the explicit construction budget; $K$ is the sampled convergence update.}\label{tab:resources}
\setlength{\tabcolsep}{4pt}
\begin{tabular}{lrrrrr}\toprule
Degree & Width & CX/macro & Depth/macro & $K$ & Total CX\\\midrule
0 & 5 & 35 & 114 & 99 & 3465\\
1 & 5 & 34 & 108 & 53 & 1802\\
2 & 5 & 34 & 108 & 36 & 1224\\
4 & 5 & 34 & 108 & 22 & 748\\
8 & 5 & 35 & 114 & 15 & 525\\
16 & 5 & 34 & 108 & 13 & 442\\
\bottomrule\end{tabular}

\end{table}

The independent controlled-baseline study completes 20 cases, records 2 timeouts and rejects 38 under its width and controlled-synthesis budgets. Its maximum completed density discrepancy is $1.437\times10^{-13}$. No rejected case is relabeled successful by substituting an easier dense transform. Input preprocessing, transform construction, transpilation and simulation times are preserved separately. The fresh campaign's generated run record and timings accompany these tables; neither simulation time nor a reduced update count is described as hardware performance.


\subsection{Chebyshev filtering: query separation and compiled cost}
\begin{figure*}[!tbp]\centering\includegraphics[width=.93\textwidth]{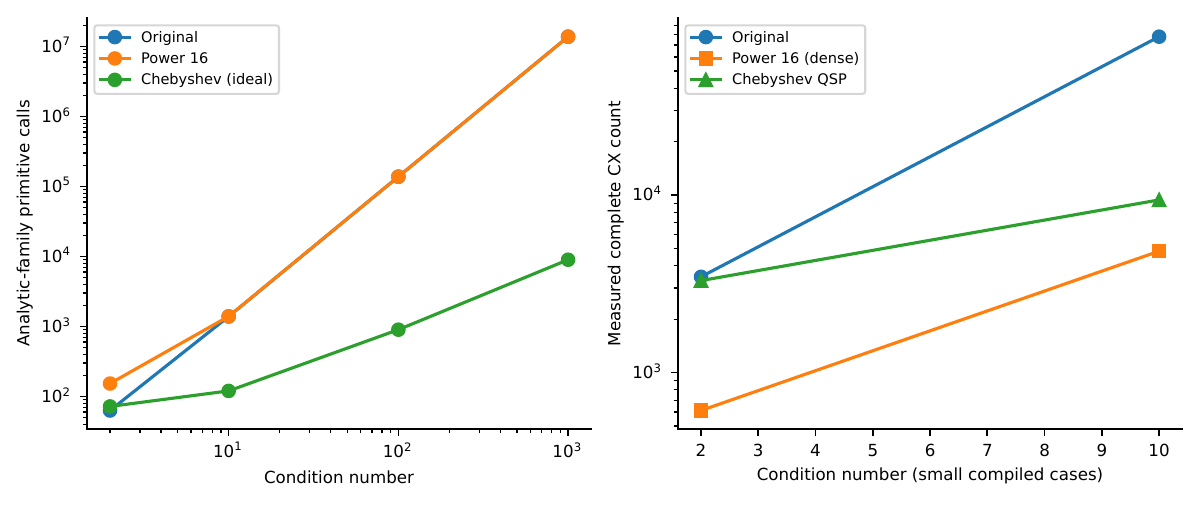}\caption{Left: numerical evaluation of the analytical slow-mode population recurrence. Right: measured CX counts of complete small native circuits under the joint accuracy and readout-reserve rule. Oracle counts and dense transformed synthesis have different costs. Neither panel measures physical hardware runtime.}\label{fig:chebcost}\end{figure*}

The distinct Chebyshev study uses the same matrices, right-hand sides, initial states and joint simulator accuracy threshold as its power comparators. Table~\ref{tab:chebstatic} reports all recorded cases. Macro caps differ: 64 for Chebyshev and the earlier condition-dependent caps for power responses. These pass counts therefore do not constitute an equal-query-budget speedup experiment. The original condition-1000 and noisy power failures remain retained.
\begin{table}[!tbp]\centering\caption{New matched study. Median charged calls is conditional on passing; the successful population changes between methods.}\label{tab:chebstatic}
\begin{tabular}{lrrr}\toprule
Method & Cases & Pass & Median charged calls\\\midrule
Original & 176 & 88 & 965\\
Power 4 & 176 & 88 & 947.5\\
Power 16 & 176 & 132 & 2363\\
Chebyshev (ideal) & 176 & 176 & 1380\\
\bottomrule\end{tabular}

\end{table}
The slow-mode family of Theorem~\ref{thm:separation} provides an independent comparison with a shared initial state and purified-distance criterion. Its exact population recurrence is numerically evaluated and cross-checked against direct small-density powers. Table~\ref{tab:cheblower} reports the resulting ideal call counts at $P\le10^{-3}$; these are analytical-family results, not compiled long circuits or a classical-inversion benchmark. Figure~\ref{fig:chebcost} contrasts them with actual small compiled gate counts.
\begin{table}[!tbp]\centering\caption{Ideal call counts for the diagonal slow-mode family, initially orthogonal to the target. Same $P\le10^{-3}$ criterion.}\label{tab:cheblower}
\begin{tabular}{lrrr}\toprule
$\kappa$ & Original & Power 16 & Chebyshev\\\midrule
2 & 63 & 153 & 72\\
10 & 1389 & 1394 & 120\\
100 & 138163 & 138176 & 900\\
1000 & 13815518 & 13815526 & 8982\\
\bottomrule\end{tabular}

\end{table}
The new circuit study completes 6 degree-budgeted native cases and 20 independent two-update full-circuit comparisons. The first are small synthesized circuits, not high-dimensional hardware runs; the second directly test reduced evolution against complete small-system circuit simulation. The maximum density discrepancy is $2.109\times10^{-14}$, and the maximum success-block discrepancy from the attenuated functional calculus is $3.331\times10^{-14}$. Larger phase degrees and oversized reset-density simulations are explicitly rejected. Phase arrays, all complete circuits and the numerical attenuation are archived.

\subsection{Measured circuits and modeled time-to-accuracy}
Table~\ref{tab:chebcost} uses complete native executions, including the stricter state-error reserve needed for the chosen readout certificate. These update counts differ from first crossings of the joint simulator criterion; both are preserved in the records. Counts assume all-to-all native synthesis. The new cost runs and the earlier power-only resource table have separate compilation records and must not be silently combined.
\begin{table*}[!tbp]\small\centering\caption{Actual complete native circuits for the common joint target plus readout-reserve rule. The final column is obtained from the complete native simulator execution; $K$ denotes macro updates.}\label{tab:chebcost}

\begin{tabular*}{\textwidth}{@{\extracolsep{\fill}}rlrrrrrr@{}}\toprule
$\kappa$ & Method & $K$ & CX & Depth & Qubits & Resets & Residual \\
\midrule
2  & Original   & 99   & 3465  & 11286  & 5 & 294  & $9.49\times10^{-7}$ \\
& Power 4    & 23   & 782   & 2484   & 5 & 66   & $2.07\times10^{-7}$ \\
& Power 16   & 18   & 612   & 1944   & 5 & 51   & $9.82\times10^{-10}$ \\
& Cheb. QSP  & 17   & 3298  & 12172  & 7 & 80   & $2.57\times10^{-9}$ \\
\midrule
10 & Original   & 2293 & 77962 & 247644 & 5 & 6876 & $5.17\times10^{-7}$ \\
& Power 4    & 461  & 15674 & 49788  & 5 & 1380 & $4.84\times10^{-7}$ \\
& Power 16   & 138  & 4830  & 15732  & 5 & 411  & $3.86\times10^{-7}$ \\
& Cheb. QSP  & 15   & 9390  & 35550  & 7 & 70   & $7.53\times10^{-11}$ \\
\bottomrule\end{tabular*}

\end{table*}
At condition 2, the original-to-QSP CX ratio is 1.05; the corresponding time ratio under the declared common duration, reuse and confidence model is 1.05. At condition 10, the original-to-QSP CX ratio is 8.3; the corresponding time ratio under the declared common duration, reuse and confidence model is 8.36. The dense degree-16 transformed compiler can be cheaper than QSP because its classical spectral transformation is outside the residual-query interface. A circuit resource saving under one compiler is therefore not a new matrix-access complexity theorem.

\begin{figure}[!tbp]\centering\includegraphics[width=.94\columnwidth]{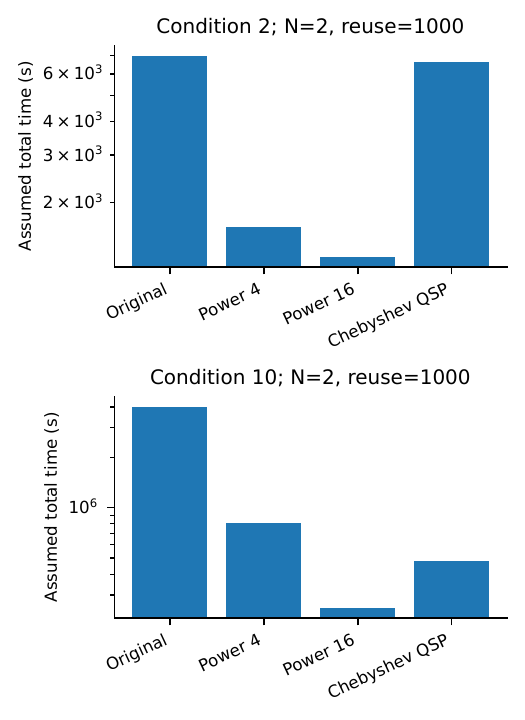}\caption{Explicit assumed-duration time model, including all confidence-dependent repeated circuit executions. These values are not measured hardware times.}\label{fig:chebtime}\end{figure}

The confidence workload uses residual target $10^{-2}$ and failure probability $0.05$, charging all three measurement settings for the two-dimensional output. The counts are conservative Hoeffding requirements calculated from an ideal state-error certificate, not measurement shots collected on a quantum device. Tighter $10^{-6}$ readout certificates are checked separately and can fail despite a small exact simulator residual. Table~\ref{tab:chebreadout} reports the conservative shot count and modeled total time for reuse 1000, reset latency $1\,\mu$s, CX duration 300 ns, one-qubit duration 30 ns and readout duration $1\,\mu$s, including synthesized measurement rotations. Figure~\ref{fig:chebtime} summarizes the resulting time model. These durations are assumptions, not a backend calibration. The very large execution budgets expose readout overhead rather than establish practical feasibility.
\begin{table}[!tbp]\centering\small\caption{Confidence-dependent execution budgets and assumed total time; $10^{-2}$ readout target. Classical preprocessing, phase synthesis and compilation are amortized over the stated reuse.}\label{tab:chebreadout}
\begin{tabular}{rlrr}\toprule
$\kappa$ & Method & $S$ & Time (s)\\\midrule
2 & Original & 6797562 & 7016\\
2 & Power 4 (dense) & 6994446 & 1627\\
2 & Power 16 (dense) & 6879777 & 1255\\
2 & Chebyshev QSP & 6986649 & 6666\\
10 & Original & 174456402 & $4.021\times10^{6}$\\
10 & Power 4 (dense) & 174457353 & $8.085\times10^{5}$\\
10 & Power 16 (dense) & 174367620 & $2.508\times10^{5}$\\
10 & Chebyshev QSP & 172410366 & $4.812\times10^{5}$\\
\bottomrule\end{tabular}

\end{table}

\subsection{Noise costs of stronger filtering}
\begin{figure}[!tbp]\centering\includegraphics[width=.94\columnwidth]{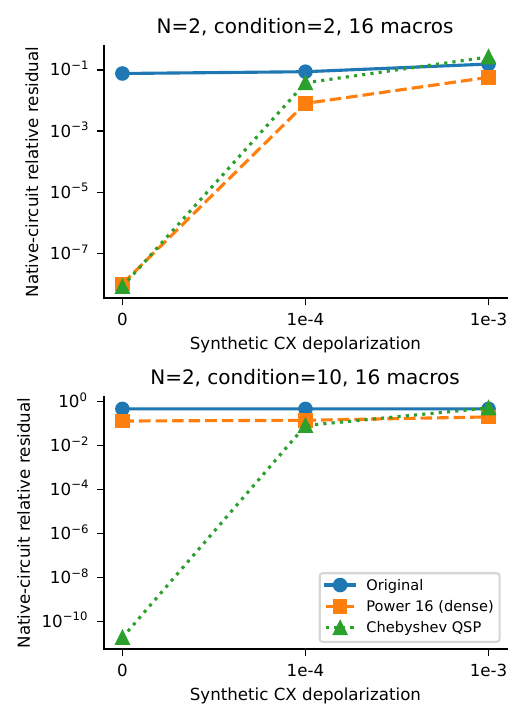}\caption{Synthetic native CX and reset noise on complete small circuits. Other gates and readout are ideal. Different macros have different gate and query costs.}\label{fig:chebnoise}\end{figure}

Figure~\ref{fig:chebnoise} compares actual native noisy simulations with the same 16-macro budget. Chebyshev macros contain more primitive calls and gates, which can worsen noise exposure even when their ideal contraction is stronger. Such a fixed-macro comparison is not an equal-gate noise benchmark. Table~\ref{tab:cheblogical} separately contrasts fixed-macro and query-coupled logical noise at high conditioning; logical improvements cannot be presented as physical gate-level evidence.
\begin{table}[!tbp]\centering\caption{New high-conditioning logical-noise study at residual target $10^{-2}$. Stationary and transient passes remain distinct from physical gate performance.}\label{tab:cheblogical}
\small\setlength{\tabcolsep}{3.2pt}
\begin{tabular}{llrrr}\toprule
Convention & Method & Cases & Transient & Stationary \\
 &  &  & pass & pass\\\midrule
fixed macro & Original & 12 & 0 & 0\\
fixed macro & Power 16 & 12 & 0 & 0\\
fixed macro & Chebyshev (ideal) & 12 & 12 & 12\\
query coupled & Original & 12 & 0 & 0\\
query coupled & Power 16 & 12 & 0 & 0\\
query coupled & Chebyshev (ideal) & 12 & 0 & 0\\
\bottomrule\end{tabular}

\end{table}
These results support a qualified contribution: a proven ideal augmented-state query separation relative to the original recurrence, a concrete small real-Julia circuit, and compiler- and model-dependent implementation savings. The native noise tests do not establish a useful large-\(\kappa\) noise threshold or a reliable readout budget on a physical backend. Calibrated-device time, robust long-circuit output and an end-to-end quantum advantage remain unmeasured.


\section{Discussion: Distinctiveness of the Recurrent Construction}\label{sec:discussion}
The primary contribution is \emph{channel design}, not the elementary polynomial $1-(1-\lambda)^{m+1}$ or the independent invention of Chebyshev eigenstate filtering. The power-polynomial channel defers resets while preserving the target kernel, and the signed-filter channel uses a QSP-compatible no-event polynomial within a continuously resident data-register recurrence. These maps define an update process, not merely a polynomial approximation applied once to an input state. Their stationary-state, tracking, error, and environment-resource results delimit this distinction.

\subsection{Direct one-shot filtering is an important comparator}
A recurring state is not required to obtain the ideal Chebyshev \emph{query} scaling for one fixed instance. In fact, the same success block $B_s=r_s(G)$ can be applied once to the reference input $\ket r$ and postselected. Writing $p_{\mathrm{dir}}=\bra r B_s^2\ket r$, kernel preservation and $q=\bra r\Pi\ket r\ge1/2$ imply
\begin{equation}\label{eq:oneshot}
 p_{\mathrm{dir}}\ge q,\qquad
 P\!\left(\frac{B_s R B_s}{p_{\mathrm{dir}}},\Pi\right)^2
 \le\frac{(1-q)\eta_s^2}{q}\le\eta_s^2.
\end{equation}
Indeed, the unnormalized target overlap of the accepted branch equals $q$, while its complementary weight is at most $(1-q)\eta_s^2$. Successful repetition has expected cost at most $1/q\le2$ under the same ideal block access. Thus a sufficiently accurate \emph{direct} Chebyshev filter also has $O(g^{-1/2}\log(1/\eps))$ ideal query scaling for a static reference start. Theorem~\ref{thm:separation} must therefore be read as a separation from the \emph{specified unfiltered original recurrence}, not from optimized direct eigenstate filtering. Equation~\eqref{eq:oneshot} is an analytical comparator, not a newly executed equal-gate experiment. The remaining distinctive motivation for recurrent operation is the retained state and its certified response to a moving target, whose conditional benefit against restarting the same channel is investigated in Sec.~\ref{sec:matchedtemporal}. A matched physical hardware-cost study remains open.

\subsection{What is demonstrated and what remains open}
The mathematical channel is valid in arbitrary fixed finite dimension under a nonzero singular gap. Implemented small circuits realize the same reduced evolution. Large simulations demonstrate selected encoded inverse actions, not universal practical success for all nonsingular matrices. Conditioning, noise floors, extraction, preprocessing, state preparation and readout can dominate. The term recurrent describes repeated resident-state channel updates; it does not establish a trainable quantum neural architecture or multi-cell physical connectivity.

Power weighting can improve macro-update counts and selected dense-compiled CX totals. The distinct Chebyshev filter changes the ideal query rate relative to this baseline, but those oracle counts exclude input preprocessing, synthesis precision, extraction and measurement. These observations are conditional on a dense transformed-unitary compiler whose classical work is substantial. The controlled-baseline construction gives an explicit oracle-style cost model, but its extra calls and controls prevent an automatic speedup conclusion. The equal-call negative result is therefore central evidence, not an incidental caveat.

The experiment coverage does not prove uniform success for every matrix of a tested order. A matrix can be arbitrarily close to singular at any order. Our preserved $\kappa=100$ noisy failures and $\kappa=1000$ finite-budget failures show exactly this boundary. Increasing degree is not a substitute for a complete physical error budget. A suitable hardware study should begin with small matched circuits, raw outcomes, reset behavior and uncertainty intervals before attempting long trajectories or large dense matrices.

The held-out cost frontier further demonstrates why a pass-rate or convergence benefit is not necessarily a resource advantage. Under destructive-readout replay and the stated weights, all 184 jointly feasible resident/fresh scenarios favor fresh preparation, although persistence alone satisfies the accuracy target in 74 additional scenarios. The appropriately restricted benefit is \emph{tracking feasibility} under a given query budget. A positive cost claim would require independently verified readout scheduling, lower resident degree/update requirements, and hardware-calibrated state-preserving operations evaluated at equal output accuracy and confidence.

\section{Conclusion}
We developed two completely positive polynomial recurrent channels for quantum solution-state preparation, including a deferred-reset power-polynomial implementation and signed Chebyshev filtering with small-instance real-Julia QSP realization. The fixed-point and moving-target results supply convergence and implementation-error bounds; the ideal query separation applies only relative to the specified unfiltered recurrence. In a 2880-case reduced-channel study, persistence improved the number of passing snapshots relative to the same restarted channel in 608 configurations, tied in 2271, and lost in one. The independent 360-scenario held-out cost study establishes the essential qualification: fresh recurrence is cheaper in all 184 cases where both recurrent methods reach the same accuracy target after destructive replay is charged; persistence alone is feasible in another 74. Direct postselected filtering is often the lowest-cost accepted-state option. The software simulations do not demonstrate finite-shot readout accuracy, noise robustness at the strict threshold, a hardware-calibrated cost advantage, or a general improvement on optimal quantum linear-system methods. The contribution is a physical recurrent-channel design and a precisely bounded conditional tracking-feasibility regime, with the resource limitations reported rather than concealed.

\section{Reproducibility and Provenance}\label{sec:provenance}
The numerical figures and tables are derived from archived records associated with the public QCeNN-MI implementation~\cite{r35}. The simulation campaigns use pinned code revisions, generated summaries, and explicit experimental settings; the source archive retains the figure files, tables, and version identifiers needed to check the reported results. The appendix-level mathematical details, historical execution caveats, and evidence-integrity audit are provided in the supplementary material.

The independent full-grid rerun of the held-out 360-scenario cost-frontier study reproduced every scenario classification, selected policy, and winning control method. Numerical differences across environments were at most $1.5\times10^{-8}$ and did not change any categorical outcome. This reproducibility evidence concerns simulated quantum channels and assumed resource weights; it is not evidence of calibrated hardware execution or finite-shot accuracy. The new rerun does not retroactively establish the provenance of earlier records with incomplete execution metadata.\par

\section*{Acknowledgment}
\phantomsection\addcontentsline{toc}{section}{Acknowledgment}
The authors gratefully acknowledge the European High Performance Computing Joint Undertaking (EuroHPC JU) for providing access to the computational resources for quantum simulation through the EuroHPC Quantum Access Pilot Call. The authors also thank the Leibniz Supercomputing Centre (LRZ) for hosting the simulation resources.



\end{document}